\documentclass{aa}  

\usepackage{natbib}
\bibpunct{(}{)}{;}{a}{}{,} 
\usepackage{hyperref}
\hypersetup{
    colorlinks=true,
    citecolor=blue,
    linkcolor=red,
    filecolor=magenta,      
    urlcolor=cyan,
    }
\usepackage[dvipsnames]{xcolor}
\usepackage{soul}
\usepackage{graphicx}
\usepackage{float}
\usepackage[caption = false]{subfig}
\usepackage[varg]{txfonts}
\usepackage{array, makecell} 
\usepackage{placeins}

\defcitealias{kravchenko18}{K18}
\defcitealias{kravchenko19}{K19}
\defcitealias{kravchenko20}{K20}
\usepackage[normalem]{ulem}

\begin{document} 

   \title{VLTI-GRAVITY measurements of cool evolved stars} \subtitle{II. Pulsation properties and mass-loss process of the Mira star R Car and the red supergiant VX Sgr} 

    \titlerunning{Pulsations and mass loss of the Mira star R Car and the red supergiant VX Sgr}
    \authorrunning{D. Jadlovský et al.}

   \author{Daniel Jadlovský\inst{1,2}
          \and
          Markus Wittkowski\inst{2}
          \and
          Andrea Chiavassa\inst{3}
          \and
          Kateryna Kravchenko\inst{4}
          \and
          Bernd Freytag\inst{5}
          \and
          Susanne Höfner\inst{5}
          \and
          Jiří Krtička\inst{1}
          \and
          Claudia Paladini\inst{6}
          \and
          Gioia Rau\inst{7,8}
          \and
          Miroslav Brož\inst{9}
          \and
          Thomas Granzer\inst{10}
          \and
          Michael Weber\inst{10}
          }

   \institute{Department of Theoretical Physics and Astrophysics, Faculty of Science, Masaryk University, Kotl\'a\v rsk\'a 2, 61137, Brno, Czech Republic  
        \and
            European Southern Observatory (ESO), Karl-Schwarzschild Str. 2, 85748, Garching bei München, Germany
        \and
            Université Côte d’Azur, Observatoire de la Côte d’Azur (OCA), CNRS, Lagrange, CS 34229, 06304, Nice Cedex 4, France
        \and
            Max Planck Institute for Extraterrestrial Physics (MPE), Giessenbachstrasse 1, 85748, Garching bei München, Germany
        \and
            Theoretical Astrophysics, Department of Physics and Astronomy, Uppsala University, Box 516, SE-751 20 Uppsala, Sweden
        \and
            European Southern Observatory (ESO), Alonso de Córdova 3107, Vitacura, Santiago, Chile
        \and
            Schmidt Sciences, New York, USA
        \and            
            National Science Foundation, 2415 Eisenhower Avenue, Alexandria, Virginia 22314, USA
        \and    
            Charles University, Faculty of Mathematics and Physics, Institute of Astronomy, V Hole\v sovi\v ck\'ach 2, 18000, Prague, Czech Republic         
        \and
            Leibniz-Institut für Astrophysik Potsdam (AIP),
            An der Sternwarte 16, 14482 Potsdam, Germany}

   \date{ }

 
  \abstract
   {}
   {The mass-loss process of red supergiant (RSG) and asymptotic giant branch (AGB) stars and its relation to variability are poorly constrained. We aim to study the photosphere and near-surface atmospheric structure, where the mass-loss is initiated. For this purpose, we studied two oxygen-rich evolved stars: the Mira-type AGB star R Car and the RSG VX Sgr.}
   {We used the VLTI-GRAVITY instrument operating in the near-infrared $K$-band. Our sample comprises 54 VLTI-GRAVITY snapshots (18 R Car, 36 VX Sgr) taken over about 7 years, making it the largest VLTI time series dataset to date. We determined the angular diameter as a function of time for the continuum (photosphere) and selected atomic and molecular bands, i.e., lines of \ion{Ti}{i} and \ion{Sc}{i} as well as bands of H$_2$O and CO. Furthermore, we compared the variability and atmospheric structure to state-of-the-art radiative-hydrodynamics CO5BOLD 3D simulations. }
   {The radii of photosphere ($R_{\star}$) and extended atmospheric layers are variable and relate to the light curve with phase shifts. The near-photospheric layers show a maximum radius near visual brightness minima ($\phi_{\rm vis} \sim 0.4-0.6 $). Inner atomic (\ion{Ti}{i}, \ion{Sc}{i}) and molecular (H$_2$O) layers are further phase-shifted by $\Delta\phi_{\rm vis} \sim 0.05 $. The more extended CO layers show longer, irregular periods and maximum extensions of $\sim 1.3-1.7 \: R_{\star}$ for R Car and of $\sim 1.5-2.2 \: R_{\star}$ for VX Sgr. Comparison with synthetic interferometric data of an AGB model based on several pulsation cycles in CO5BOLD simulations revealed a similar behavior. The photosphere shows regular pulsations, but with maximum diameters preceding minimum brightness ($\phi_{\rm vis} < 0.5 $). The H$_2$O layer showed a much weaker extension compared to our observations, while CO showed a good agreement. Furthermore, during the 2020-2021 season, VX Sgr exhibited an extreme mass-loss event similar to that of Betelgeuse, preceded by two strong shocks and culminating with the extreme expansion of H$_2$O and CO layers, both up to $\sim 2.2 \: R_{\star}$. Unexpectedly, during this event, we also detected Brackett $\gamma$ in interferometric data as well as strong Balmer emission in optical spectra, both of which are also signatures of a shock propagating through the atmosphere.}
   {The Mira R Car showed an estimated photospheric radius of $R_{\star} = 280 \pm 25 \: \rm R_\odot$, with a regular fundamental mode (FM) pulsation amplitude of $\sim13 \%$ of $R_{\star} $. During its active cycle, the extreme RSG VX Sgr showed $R_{\star} = 1556 \pm 110 \: \rm R_\odot$, with an FM pulsation amplitude of $ \sim13 \%$ of $R_{\star} $, the same as R Car. During its quiescent cycle, it showed a smaller value, $R_{\star}= 1456 \pm 108 \: \rm R_\odot$, and low-amplitude pulsations near the first overtone (O1), only  $\sim4 \%$ of $R_{\star} $. This supports a steady mass-loss process for Mira stars related to stable large-amplitude FM pulsation, whereas the mass-loss process for RSGs may be dominated by extreme events connected to changes in the pulsation mode from low-amplitude O1 to large-amplitude FM pulsations.} 

   
   \maketitle
%

\section{Introduction}
Red supergiants (RSGs) and red giants on the asymptotic giant branch (AGB) experience significant mass loss and are thus some of the main contributors to dust production in the interstellar medium \citep[e.g.,][]{tielens05, van_loon05, levesque17, chiavassa24}. From the evolutionary point of view, these groups of evolved stars are quite different: AGBs are low- and intermediate-mass stars, whereas RSGs are massive stars that will evolve toward core-collapse supernovae (SNe). On the other hand, from the observational point of view, these groups of evolved stars share many similar characteristics, such as variability, high mass-loss rates, and extended atmospheres \citep[e.g.,][]{arroyo15, chiavassa24}. The mass-loss rates range from $ 4 \times 10^{-8}$ to $ 8 \times 10^{-5} \: \rm  M_{\odot} \, yr^{-1}$ for the Mira-type AGB stars, while for the more massive RSGs, the range is from $2 \times 10^{-7} $ to $ 3 \times 10^{-4} \: \rm  M_{\odot} \, yr^{-1}$ \citep[e.g.,][]{debeck10}. Despite the importance of the mass loss from the cool evolved stars to the chemical evolution of the Universe, we still do not have a comprehensive understanding of all the physical processes that come into play \citep{chiavassa24}, especially for the RSGs.

For Mira-type AGB stars, the mass-loss mechanism is better understood. They are radial fundamental-mode pulsators \citep{wood99, wood00} with regular large-amplitude changes in brightness, up to several magnitudes in $V$ band. In this case, the mass loss triggered by pulsation and convection would be the most effective. Pulsation and large-scale convective flows levitate the material to a distance far enough from the star where the cooler extended atmosphere allows for molecule and dust formation. Subsequently, the radiation pressure on dust grains becomes effective and drags the gas, driving mass loss \citep{woitke06, hofner18, bladh19, freytag23}. For carbon-rich (C/O $>1$) AGB stars, the wind is driven by amorphous carbon, whereas for the oxygen-rich (C/O $<1$) stars, the wind is driven by silicate grains \citep{hofner22}. Dynamic 1D CODEX self-excited pulsation models for Mira variables have shown that pulsations of the photosphere are quite regular and generate global shocks, but with increasing distance, the variability of extended layers becomes more irregular \citep{ireland08, ireland11}. However, 3D CO5BOLD radiative hydrodynamics (RHD) simulations give us a more complex picture. While the radial pulsations of the star are present, the surface is also covered by various large-scale non-radial structures, such as the giant convective cells. The interaction between convection and pulsation generates large-scale shocks, resulting in increasingly patchy and clumpy structures in the extended atmosphere and the dust formation zone \citep{frey17, hofner19, freytag23}. Observationally, the presence of extended atmospheres up to several stellar radii ($ R_{\star}$) have been confirmed for AGBs \citep[e.g.,][]{mennesson02, ohnaka05, wittkowski18} as well as various large-scale structures on the surface \citep[e.g.,][]{paladini18, vlemmings24}.

The same mechanism appears less effective for RSGs (as well as for semi-regular AGB stars). There, the variability is more complex, consisting of several modes, such as pulsations in higher overtones and long secondary periods \citep[LSPs;][]{wood99, kiss06, wood09}, resulting in smaller brightness amplitudes and irregular variability. Recently, it has been shown that LSP photometric variations may be caused by low-mass companions \citep{decin25}, as evidenced by Betelgeuse \citep{goldberg24, macleod25, howell25, dupree26}. Based on 1D and 3D dynamical models of RSGs, \citet{arroyo15} and \citet{wittkowski16} suggested that pulsation and convection in these stars, based on current models, are not effective enough to lift the atmosphere to such a distance where dust could form. Nonetheless, comparable extended atmospheres and large-scale surface structures similar to those of AGBs were also observationally confirmed for RSGs using interferometry \citep[e.g.,][]{haubois09, chiavassa10a, arroyo13, wittkowski17, climent20, chiavassa22, gonzalez24}, while the extended atmospheres of RSGs were also recently directly imaged by \citet{guarcello24} using JWST. Narrow line emission components in the first days of a Type II SN explosion have also further proven the existence of a dense circumstellar medium (CSM) around RSGs \citep[e.g.,][]{franson14} as well as of enhanced mass loss of RSGs before the SN \citep[e.g.,][]{jacobson22}. Additionally, empirical extended atmospheres added to 1D MARCS models have shown good agreement with interferometric observations \citep{davies21, gonzalez23, gonzalez24}. Taken together, these findings indicate a missing physics in our current understanding of the RSG winds, namely, how the mass-loss process is triggered in the near-surface regions and how the material is levitated to form the observed extended atmospheres. Alternative mass-loss mechanisms have been proposed, such as radiative pressure on molecular lines \citep{josselin07}. The recent event of the Great Dimming of Betelgeuse \citep{guinan20, levesque20, dupree20, dharma20, montarg21, taniguchi22} indicates that episodic mass-loss events could be the missing component to our understanding the mass{-}loss process of RSGs \citep{dupree22, humphreys22}. In particular, powerful shocks propagating through the photosphere on a timescale of a few weeks were associated with this mass ejection \citep{kravchenko21, jadlovsky24}. Recently, \citet{freytag24} demonstrated, using their CO5BOLD simulations, that RSGs may irregularly show dimming events due to obscuration by patches of cool gas, while for AGBs, this behavior is characteristic of their pulsation cycles.

To date, not many interferometric studies have been able to analyze the temporal variability of extended atmospheric layers of cool evolved stars. \citet{thompson02} was one of the first to analyze the near-infrared (IR) variability of the extended layers of two Mira variables, RZ Peg and S Lac, using the Palomar Testbed Interferometer (PTI). Later, \citet{thompson03} presented results for many more Mira stars and found different pulsation properties for oxygen and carbon-rich stars. The full PTI sample consists of about 100 Mira stars, observed during 1998-2008. Recently, analysis of the full sample (i.e., the Mira Reference Dataset) was resumed by \citet{aguirre24}, who reported systematic phase shifts of the continuum and molecular bands with respect to the visual light curve. Meanwhile, \citet[][Paper I]{wittkowski18} used the VLTI and the GRAVITY instrument to study the variability of the atmospheric extension for a Mira-type star R Peg over one pulsation cycle and found that the maximum radius of continuum and H$_2$O layers were anticorrelated with the light curve, while the CO layers were correlated. \citet{rosales23, rosales24} studied the variability of surface features of R Car and found that substructures are correlated with the pulsation phase. Until now, no similar study was done for RSGs, apart from \citet{GRAVITY21}. They analyzed GCIRS 7, a RSG in the Galactic center, and demonstrated that RSGs also show significant diameter variations of the photosphere and molecular layers.

In this work, we analyze the largest sample of spectro-interferometric VLTI-GRAVITY observations of cool evolved stars to date. We study the relation of the extended atmosphere to the pulsation cycle. We aim at comparing a time series of the spatially resolved variability of an AGB and a RSG star to better understand the similarities and differences between cool evolved stars of different types at the onset of the mass-loss process close to the stellar surface. We also compare these observations to the most recent state-of-the-art 3D dynamical models \citep{ahmad23}. The first object is the oxygen-rich AGB star R Car, which is a typical M-type Mira variable, with strong single-mode pulsations and amplitude of brightness variations of $\Delta m_V \: \rm  \sim 7.4 \: \rm mag$ \citep{rosales24}. The second object is 
VX Sgr, an extreme RSG exhibiting irregular variability, with a longer period during active cycles and a shorter period during quiescent cycles \citep{tabernero21}. While its large peak-to-peak amplitude of brightness variations up to $\Delta m_V \: \rm   \sim 6 \: \rm mag$ during its active cycles resemble a Mira variability \citep{lockwood82}, 
it is considered to be a massive supergiant experiencing an extensive mass loss \citep{gail20}. However, it has also been hypothesized to be a super-AGB star or Thorne–Żytkow object \citep[T$\rm \dot{Z}$O,][]{tabernero21, ogrady26}. In Table~\ref{table:table_params}, we list the fundamental parameters of our objects from the literature. Neither of our targets is reported to show the LSPs.

The present paper is structured as follows. In Sect.~\ref{chapter:observations} we describe our dataset and data reduction methods. In Sect.~\ref{chapter:res} we present our results on the variability of the photosphere and extended atmosphere, whereas in Sect.~\ref{chapter:models} we compare this to similar features from state-of-the-art 3D RHD dynamical models. In Sect.~\ref{chapter:discussion} we discuss the implications of our results, and then in Sect.~\ref{chapter:conclusions} we summarize our findings.


\begin{table}[t]
        \centering
        \caption{Fundamental physical parameters.} 
        \label{table:table_params} 
        \setlength{\extrarowheight}{1pt}
        \begin{tabular}{cccc}
\hline\hline
\text{Parameter} & \text{R Car (Mira)}  & \text{VX Sgr (RSG)} & \text{References}    
\\
\hline
$ \text{Sp. type}  $ & M5-M8  & M3-M8  &  \\
$ \Delta m_V \: \rm [mag]  $ & $\sim$7.4  & $\sim$6.0 & AAVSO  \\
$ m_K \: \rm [mag]  $ & -1.23 & -0.17 & 5 \\
$P \: \rm [d]$ & 304-314 & 754$\pm 56$ & 1, 2, 5 - 11 \\ 
$d \: \rm  [pc]$ & 182 $ \pm 16 $ & $1560 \pm 110 $ & 6 - 12  \\ 
$T_{\rm eff} \: \rm [K] $ & 2800-3100  & 2900-3700  & 3, 7 - 9, 10\\ 
$ \log(L/\rm L_\odot)$ & $\sim 3.7$  & $ \sim 5.3$   & 4,7 - 12 \\ 
$R_{\star} \: \rm [R_{\odot}] $ & 240$^{+38}_{-33}$  & 1350–1940 & 7 - 10 \\ 
$M \: \rm [M_{\odot}]$ & 0.87$^{+0.47}_{-0.31}$  & 10-40 & 7 - 9, 14 \\
$\dot{M} \: \rm [M_{\odot} \, yr^{-1}]$ & $ 1.6 \times 10^{-9}$  & $ 2 \times 10^{-5}$  & 4, 8 - 13 \\
\hline
        
        \hline      
        \end{tabular}
\tablefoot{AAVSO - American Association of Variable Star Observers, 1 - \citet{samus17}, 2 - \citet{vogt16}, 3 - \citet{mcdonald12}, 4 - \citet{groenwegen99}, 5 - \citet{whitelock08}, 6 - \citet{gaia21}, 7 -\citet{takeuti13}, 8 - \citet{scicluna22}, 9  - \citet{tabernero21}, 10 - \citet{lockwood82}, 11 - \citet{kiss06}, 12 - \citet{xu18}, 13 - \citet{mauron11}, 14 - \citet{arroyo15} }  
\end{table}

\section{Observations and data reduction}
\label{chapter:observations}
\subsection{Observations}

We used near-IR spectro-interferometric observations from 2018 to 2025 for our two targets, obtained with the VLTI-GRAVITY instrument \citep{GRAVITY17}, namely 18 observations for R Car and 36 for VX Sgr. The instrument operates in the near-IR $K$-band ($2.0 - 2.4 \: \rm \mu m $) and we employed the high spectral-resolution mode ($R \sim 4000$). Tables \ref{table:table_vxsgr} and \ref{table:table_rcar} list the VLTI observations of R Car and VX Sgr, respectively. A major part of the data originates from our time series monitoring programs (105.207Y, PI: Wittkowski and 115.27VK, PI: Jadlovsk\'y), and the remaining data consists of archival data from other programs (0100.D-0835, PI: Sanchez, 0101.D-0616, PI: Wittkowski, 0102.D-0197, PI: Wittkowski, and 0103.D-0245, PI: Kravchenko).

For each of our science observations, we took two observations of an interferometric calibrator (CAL-SCI-CAL sequence), which improves the quality of visibility calibration. The majority of the older archival data were taken in the CAL-SCI sequence, and therefore they had only one calibrator. The details about the calibrators are listed in Table~\ref{table:table_cals}. As our targets are bright, all the data were taken in split polarization mode to increase the internal fringe contrast.

For all the observations, we used the small configuration (stations A0-B2-D0-C1) of the four Auxiliary Telescopes (ATs), resulting in baseline lengths between about $ 11.3 \: \rm m $ and $ 33.9 \: \rm m $. The choice of the VLTI-AT configuration was driven by our observing strategy, which is focused on the variability of our targets at scales of the overall stellar diameter. For this purpose, the most suitable are spatial frequencies within the first lobe of the visibility function. Since our targets have a large angular diameter of $ \sim 10 \: \rm mas $, the small configuration allows us to study the relevant spatial frequencies. The higher spatial frequencies (larger configurations) would be sensitive to substructures of the stellar disk, which is not the focus of this work.

\subsection{Data reduction}
We downloaded the raw data from the ESO archive and processed it using the \textit{ESO Reflex} workflow for VLTI-GRAVITY in its version 1.6.7\footnote{\url{https://www.eso.org/sci/software/pipelines/gravity/gravity-pipe-recipes.html}}.
First, we used the \textit{gravity\_wkf} workflow to reduce the raw data of our science target and its calibrator(s). The calibrators were chosen as stars with a well-known angular diameter (based on \citealt{bourges17}; see Table~\ref{table:table_cals}) and thus well-known visibility, allowing us to calibrate the visibility of our science targets. Our targets require short exposures (as they are bright), allowing for several objects and sky observations to be taken as part of a single observing block. Therefore, apart from the default settings in \textit{gravity\_wkf}, we also used \textit{average-vis} and \textit{average-sky} to average object and sky observations. 

For the reduced data, we used \textit{gravity\_viscal} to calibrate the visibility of the science target using the visibility of the calibrator(s). This allowed us to obtain the final calibrated science visibility. 
The formal visibility errors included in the data are small ($\sigma_{\rm rel} \lesssim 0.002$). For each observation with 2 calibrators ($\rm CAL_1$ and $\rm CAL_2$), we estimated realistic uncertainties by determining the relative error from the difference in the visibility transfer function of the calibrators as $\rm (|TF_1 - TF_2|)/ (TF_1+TF_2)$, yielding $\sigma_{\rm rel} \sim 0.01-0.05$ for different observations. The errors were quadratically added to the errors in the original data. For observations taken with only one calibrator, we used the average error value from the epochs with 2 calibrators. The split polarization mode produces two sets of visibility data for each baseline, which we averaged for the final analysis. Fringe-tracking was performed on-axis; thus, the dataset also includes the low-resolution observations of the science target from the fringe tracker (FT). 

We calibrated the wavelength scale in our data using a telluric model produced with the SkyCalc tool\footnote{\url{https://www.eso.org/observing/etc/bin/gen/form?INS.MODE=swspectr+INS.NAME=SKYCALC}} \citep{noll12, jones13}. This resulted in an average shift of $ \sim -11 \rm \:  \AA$ for the observations taken before the GRAVITY spectral grism change in October 2019, and a shift of $ \sim -6.5 \rm \: \AA$ for the newer data. Then, we also calibrated the flux. We used the MARCS stellar atmosphere models \citep{gustaf08} to calibrate the flux of calibrators. The obtained spectral transfer function was used to calibrate the flux of our science targets. The parameters of synthetic MARCS spectra used for each calibrator are listed in Table~\ref{table:table_cals}. All calibrated spectro-interferometric data are plotted in Appendices \ref{appendix:r_car_data} and \ref{appendix:vx_sgr_data}, including the FT data. In all the data, the FT data agree with the science data, confirming the high accuracy of the absolute visibility calibration.

\subsection{Complementary data}
We used visual and $V$ light curves of R Car and VX Sgr from the American Association of Variable Star Observers (AAVSO\footnote{\url{https://www.aavso.org}}). This allowed us to establish the period phase for each epoch (with $\phi_{\rm vis} = 0$ at maximum visible brightness). Additionally, to further discuss the unusual variability of VX Sgr (see Sect. \ref{chapter:discussion}), we also compared it to optical spectra from the STELLA telescope \citep{stella, stella_2, stella_3}. We used the radial velocity data from STELLA, namely observations from 2016 to 2021 (PI: Dorda) and from 2024 to 2025 (PI: Jadlovsk\'y). 

Lastly, for converting the determined diameters from angular units to physical size, for R Car, we adopt the distance $d$ of $182 \pm 16 \: \rm pc $ \citep{gaia21}. For VX Sgr, distances determined based on \textit{Gaia} measurements have a large uncertainty. Fortunately, VX Sgr is a strong maser source; therefore, there are more precise measurements of distance based on masers available. Namely, $1.57 \pm 0.27 \: \rm kpc $ \citep{chen07} or $1.56\pm 0.11  \: \rm kpc$ \citep{xu18}. We adopted the latter more precise value.

\begin{figure}[t]
    \includegraphics[width=0.5\textwidth, keepaspectratio]{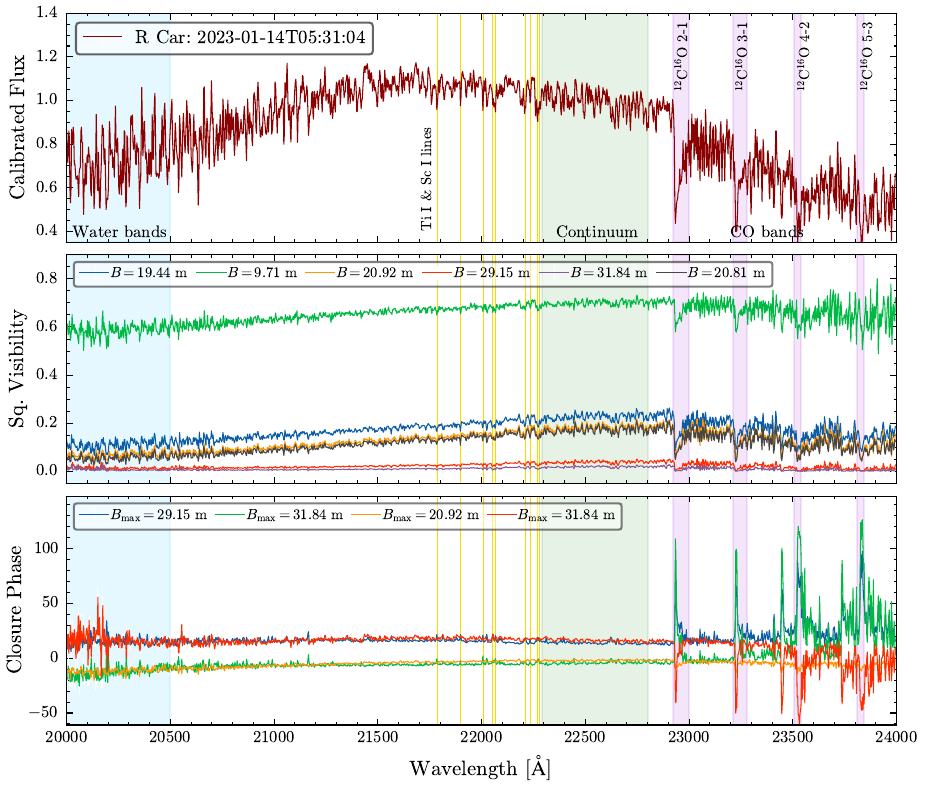}
    \includegraphics[width=0.5\textwidth, keepaspectratio]{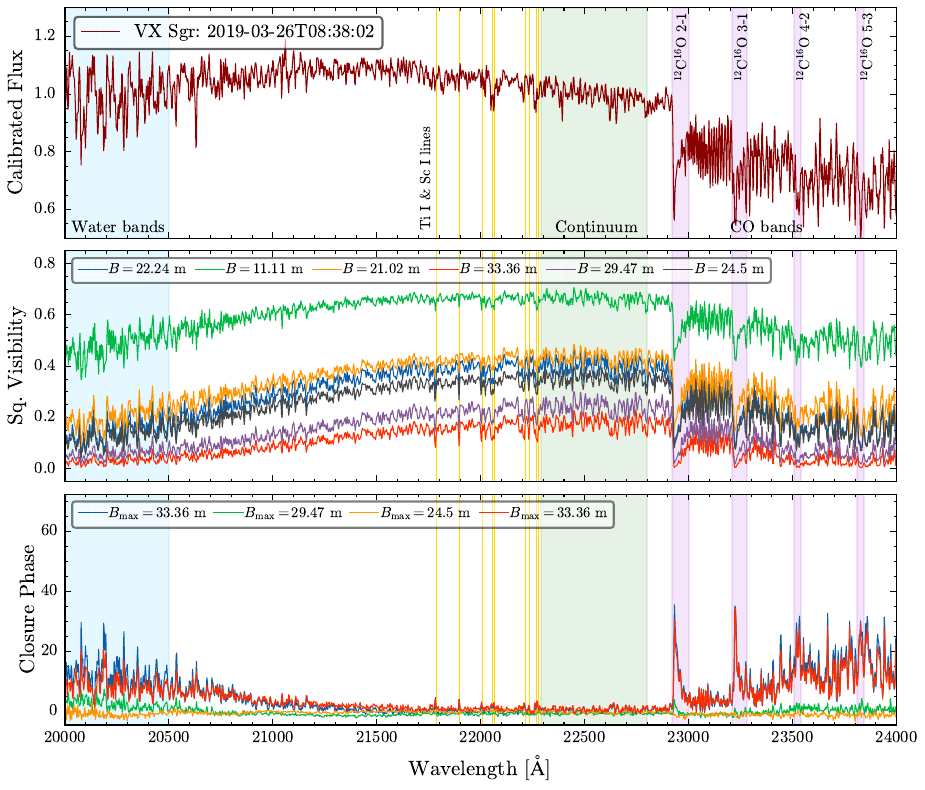}
    \caption{Examples of the calibrated flux (averaged for four ATs), the squared visibility amplitude $|V|^2$ (averaged for both polarizations), and the closure phase as a function of the wavelength for both our targets at representative dates. The flux was calibrated using the spectral transfer function of the calibrators. The $|V|^2$ dataset consists of six projected baselines $B$, while the closure phase consists of four baseline triplets, labeled using the longest baseline in the triplet $B_{\rm max}$. Spectral regions analyzed in this work are highlighted. 
    {\em Upper three panels:}
    R Car.
    {\em Lower three panels:}
    VX Sgr.
    }
    \label{fig:spectrum}
\end{figure}

\section{Results}
\label{chapter:res}
\subsection{Qualitative description}
\begin{figure*}[t]
    \includegraphics[width=1\textwidth, keepaspectratio]{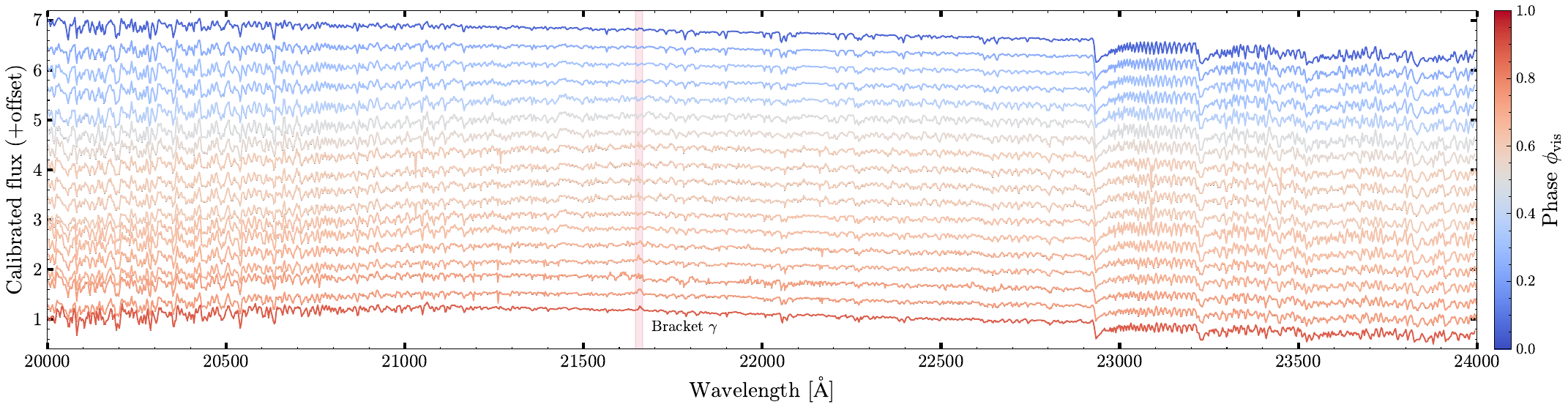}
    \includegraphics[width=1\textwidth, keepaspectratio]{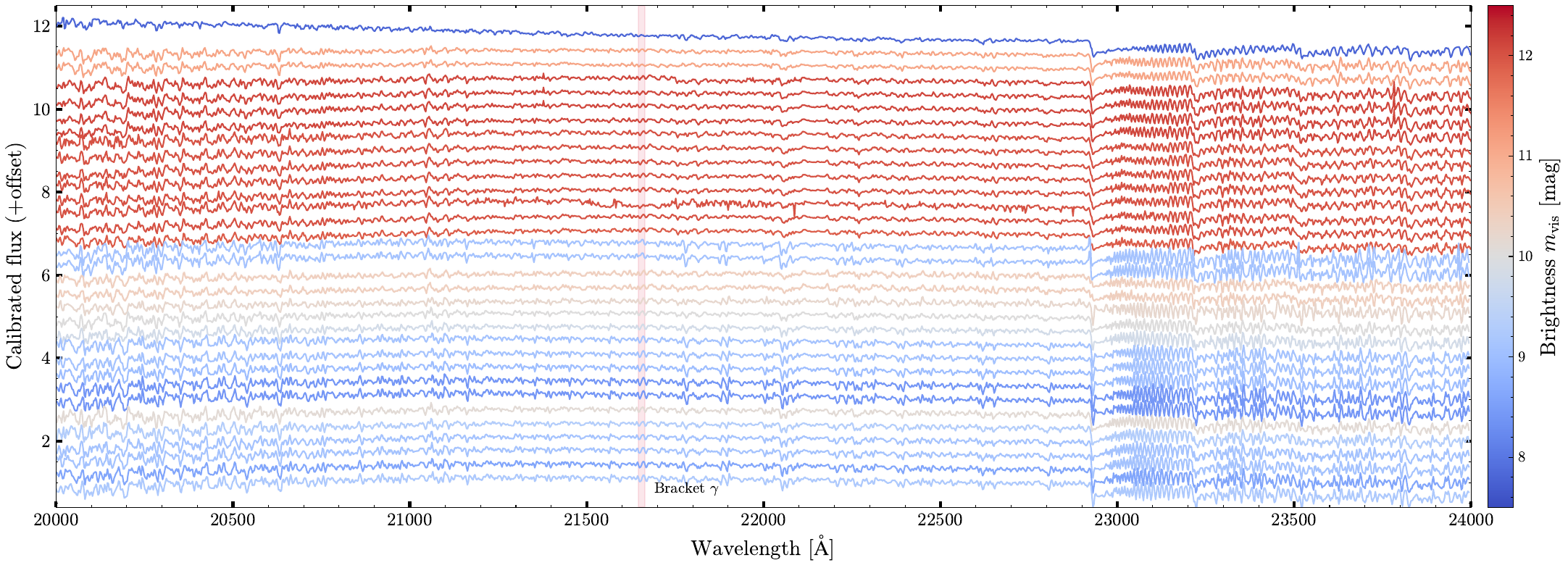}
    \caption{Variability of the flux (calibrated using the spectral transfer function) for all our observations. The spectra were divided by mean flux in the continuum region ($2.22-2.28 \rm \:\mu m $). 
    {\em Upper panel:} R Car. The spectra were vertically shifted, ordered, and color-coded by phase $\phi_{\rm vis}$ ($\phi_{\rm vis} =0$ corresponds to brightness maximum). 
    {\em Lower panel:} Same, but for VX Sgr.  The spectra were vertically shifted, ordered by observation date, and color-coded by the nearest $m_{\rm vis}$ values from AAVSO.} 
    \label{fig:spectrum_full}
\end{figure*}

\begin{figure}[t]
    \includegraphics[width=0.5\textwidth, keepaspectratio]{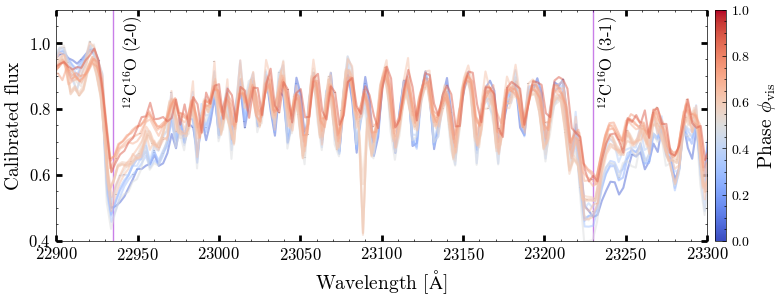}
    \includegraphics[width=0.5\textwidth, keepaspectratio]{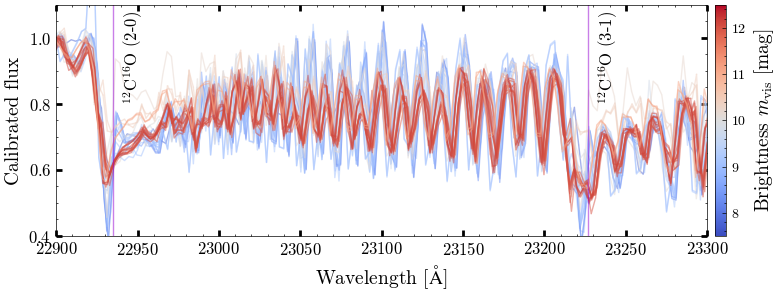}
    \caption{Same as Fig. \ref{fig:spectrum_full} but zoomed in on the CO molecular bands (2-0, 3-1). The spectra show deeper CO absorption near the brightness maxima.
    {\em Upper panel:} R Car.
    {\em Lower panel:} VX Sgr.}
    \label{fig:spectrum_co}
\end{figure}

We begin with a qualitative description of the interferometric dataset. Figure \ref{fig:spectrum} shows an example of a typical calibrated flux, squared visibility amplitude ($ |V|^2$), and closure phases 
plotted versus wavelength for each of our two science targets. The first and fourth panels of Fig.~\ref{fig:spectrum} show that the calibrated flux of both our targets, the Mira-variable R Car and the RSG VX Sgr, are similar and typical of other oxygen-rich AGBs and RSGs \citep[e.g.,][]{lancon00, lancon07}. Both our targets have the peak flux near the continuum-dominated middle of the passband at $ 2.2 \: \rm \mu m $. The flux decreases toward both sides of the passband, below $ 2.1 \: \rm \mu m $ due to the absorption bands of H$_2$O, and above $ 2.3 \: \rm \mu m $ due to the strong absorption bands of CO, as well as weaker bands of H$_2$O. In Figs.~\ref{fig:spectrum_full} and \ref{fig:spectrum_co}, we show the flux variability, for the full spectrum, as well as for molecular bands of CO. For R Car, we see anticorrelated behavior for CO and H$_2$O - the CO absorption is the strongest during the brightness maximum ($\phi_{\rm vis} \sim 0$), while it is the weakest for H$_2$O, sometimes with peak flux shifted to $ \sim 2.1 \: \rm \mu m $ (due to the higher $T_{\rm eff}$, possibly indicating thermal destruction of H$_2$O molecules near the maximum brightness). The opposite situation is reached during the brightness minimum ($ \phi_{\rm vis} \sim 0.5$), when the absorption in H$_2$O becomes stronger. A similar behavior can be seen for VX Sgr, but it shows more complicated variability. The spectra of R Car show the emergence of hydrogen Brackett $\gamma$ (Br$\gamma$) emission ($2.167 \rm \: \mu m$) before the brightness maximum ($ \phi_{\rm vis} = 0.7-1.0$). For some Miras, hydrogen emission is known to appear as part of their pulsation cycle before the maximum brightness, due to propagation of shocks through the atmosphere, although it is usually observed for Balmer and Paschen series \citep[e.g.,][]{fox84, lancon00, fabas11}.

Likewise, the squared visibility amplitude ($|V|^2 $) in the second and fifth panel of Fig.~\ref{fig:spectrum} shows similar features as observed for other oxygen-rich Mira-variables and RSG stars \citep[e.g.,][]{perrin04,arroyo15, wittkowski18, gonzalez24}. There is a maximum of $|V|^2$ near the continuum-dominated region $ \sim 2.2 \: \rm \mu m $, which would correspond to the smallest angular diameters, i.e., the closest to the photosphere. $|V|^2$ drops to lower values at wavelengths corresponding to H$_2$O bands and even more significantly at CO bands. VX Sgr also shows more prominent atomic features extended with respect to the continuum, as has been shown by, for example, \citet{chiavassa10a}. Brackett $\gamma$ is not detected in interferometric data for R Car.

In the third and sixth panel of Fig.~\ref{fig:spectrum}, we also show closure phases, even though they are not directly used in our analysis, and we discuss them only qualitatively. They show quite significant variations at most dates, at wavelengths corresponding to H$_2$O bands, but especially at those of CO bands. This could suggest substructures within the overall structure or asymmetries due to spatially localized outflows. Nonetheless, this is primarily caused by the fact that at some baselines in the triplet the star is already quite resolved. Indeed, by comparing the longest baseline in the triplet ($B_{\rm max}$) for VX Sgr in the sixth panel of Fig.~\ref{fig:spectrum}, the two triplets with shorter $B_{\rm max}$ do not show a large deviation from $0 \: \rm \deg$. For a few dates, the molecular bands also became more resolved and reached the first null point of the visibility function at the longest baselines, and thus the closure phase changed from $ 0 \deg $ to $ 180 \deg $, while the signal of possible asymmetries or substructures became enhanced.

\subsection{Spectral selection of the squared visibilities}
The main objective of our work is to study the overall variability of the atmospheric extension of distinct spectral features in the $|V|^2$ data. In the following subsections, we describe in more detail the selected atomic and molecular features, and we give their overview in Table~\ref{table:table_lines}. 

A similar approach was used in several previous works, e.g., \citet{wittkowski18,arroyo15,gonzalez24}, while probably the most detailed work was done by \citet[][hereafter \citetalias{kravchenko20}]{kravchenko20}. They used the tomographic method to study lines forming in given ranges of optical depths in the atmosphere. They used the VLTI-AMBER instrument \citep{amber} in its high-resolution mode ($R \sim 12 \, 000$), which had three times the resolving power of VLTI-GRAVITY ($R \sim 4000$). Therefore, they could also resolve weaker atomic lines than in this work.

\subsubsection{Continuum}
To study the variability of the photosphere, we selected the region of the $K$ band without strong atomic lines or molecular bands, dominated primarily by the continuum. The region most suitable for this is between $2.22$ and $2.28 \: \rm \mu m $. \citetalias{kravchenko20} showed that the strongest features contributing in this region consisted of \ion{Ca}{i}, \ion{Si}{i}, and \ion{Fe}{i} atomic lines and weak molecular bands of CN, all corresponding to the innermost layer of the atmosphere. However, the resolution of VLTI-GRAVITY is not sufficient to resolve these atomic and molecular features. Therefore, to calculate the angular diameter of the continuum $ \theta_{ \rm cont} $, which corresponds to the stellar photosphere, we make an average of $|V|^2$ in this region, using only $|V|^2$ larger than its median as an arbitrary threshold to remove contributions by slightly extended atomic and molecular features. 

\subsubsection{\ion{Titanium}{I} and \ion{Scandium}{I} atomic lines}
The strongest atomic features in our data are the lines of \ion{Ti}{i} and \ion{Sc}{i}. These lines belong to an intermediate layer of atmosphere in \citetalias{kravchenko20}, forming above the continuum. Indeed, $|V|^2$ in these lines significantly drops compared to the continuum, especially for VX Sgr. Therefore, we analyzed these lines separately from the continuum. To determine the angular diameter of these atomic layers $ \theta_{ \rm \ion{Ti}{i}, \ion{Sc}{i}} $, we take the minimum $|V|^2$ value of each line. 

\subsubsection{Water molecular bands}
The part of the spectrum that is the most dominated by the H$_2$O molecular bands is located between $1.97$ and $2.05 \: \rm \mu m $. However, it consists of many weaker H$_2$O bands, and therefore, the resolution is not sufficient to determine specific band heads. Thus, to determine the angular diameter $ \theta_{ \rm H_2O} $ in this region, we make an average of $|V|^2$ values lower than the median as an arbitrary threshold. Some H$_2$O bands were also part of the intermediate atmospheric layer by \citetalias{kravchenko20}, but from our data, it is clear that the H$_2$O molecular bands form above the atomic layer of \ion{Ti}{i} and \ion{Sc}{i}, and therefore we analyzed these features separately. Some H$_2$O molecular bands also form above $>2.3 \: \rm \mu m $, monotonically decreasing the visibility function in between the CO band heads. In this work, we focus only on the H$_2$O bands below $<2.05 \: \rm \mu m $.

\subsubsection{Carbon monoxide molecular bands}
The most prominent features in $|V|^2$ data are the CO molecular bands, which consistently represent the most extended layers. The same was the case in \citetalias{kravchenko20}, where their most extended layer consisted solely of CO bands. We can resolve specific molecular band heads of $^{12}$C$^{16}$O corresponding to vibrational transitions $v= 2-0 \: (\text{\rm first overtone}), 3-1, 4-2 $, and $ 5-3$. We can also resolve the overtones for the isotopologs $^{13}$C$^{16}$O and $^{12}$C$^{18}$O, while the overtone of the isotopolog $^{12}$C$^{17}$O is blended with the $v= 3-1$ transition of $^{12}$C$^{16}$O. We focus on the transitions of $^{12}$C$^{16}$O, which are the strongest features. To determine the angular diameters $ \theta_{ \rm CO (2-0)}, \theta_{ \rm CO (3-1)}, \theta_{ \rm CO (4-2)},$ and $ \theta_{ \rm CO (5-3)}$, we use the minimum $|V|^2$ of each band head, i.e., the maximum extension. 

\begin{table}[t]
        \centering
        \caption{Wavelength regions used for our analysis. The reference optical depth, $\tau_0$, was computed at $\lambda = 5000 \: \rm \AA$ from \citetalias{kravchenko20}.} 
        \label{table:table_lines} 
        \setlength{\extrarowheight}{1pt}
        \begin{tabular}{ccc}
\hline\hline
\text{Region} & \text{Wavelength [$\mu \rm m $]} & \text{$\log \tau_0$} \\
\hline
Continuum & 2.22-2.28 & $-0.65<\log \tau_0 \lesssim 0.25^{c}$ \\ 
$\ion{Ti}{i}$ and $ \ion{Sc}{i}$ & 2.18-2.22$^a$ & $-1.75<\log \tau_0 < -1.00$ \\ 
H$_2$O & 1.98-2.05 & $-1.75<\log \tau_0 < -1.00$ \\ 
$^{12}$C$^{16}$O (2-0) & 2.29$^{b}$ & $-3.00^{d}<\log \tau_0 < -2.10$  \\ 
$^{12}$C$^{16}$O (3-1) & 2.32$^{b}$ & $-3.00^{d}<\log \tau_0 < -2.10$ \\ 
$^{12}$C$^{16}$O (4-2) & 2.35$^{b}$ & $-3.00^{d}<\log \tau_0 < -2.10$ \\ 
$^{12}$C$^{16}$O (5-3) & 2.38$^{b}$ & $-3.00^{d}<\log \tau_0 < -2.10$ \\ 
        \hline
        \end{tabular}
\tablefoot{ 
\small $^a$ Only lines in the indicated region were selected. \small $^{b}$ Minimum $|V|^2$ value in the band head. \small $^{c}$ The optical depth range corresponds to weak unresolved lines from the innermost layer from \citetalias{kravchenko20}, which may contaminate our continuum. \small $^{d}$ The maximum extent of the atmospheric model in \citetalias{kravchenko20}.} 
\end{table}

\begin{table*}[htbp]
        \centering
        \caption{Results related to the mean angular diameters and phase shifts for the analyzed spectral regions. The parameters of the sine fit are listed as well as the average values for layers with no fit.} 
        \label{table:table_results} 
        \setlength{\extrarowheight}{1pt}
        \begin{tabular}{cccccc}

\multicolumn{5}{c}{R Car} \\ 
\hline\hline
\text{Region} & \text{Mean angular diameter $ \theta $ [mas]}  & \text{Scaling factor $A$} & \text{Phase shift $ \Delta \phi_{\rm vis}$}   & \text{Amplitude $\Delta \theta $ [mas]}  \\ 
\hline
Continuum & 14.33$_{\pm0.11}$ & 0.92 & 0.43$_{\pm0.01}$  & 1.81 \\ 
$\ion{Ti}{i}$ and $ \ion{Sc}{i}$ & 16.21$_{\pm0.11}$ & 0.92 & 0.46$_{\pm0.01}$  & 1.51 \\ 
H$_2$O & 17.92$_{\pm0.34}$ & 0.86 & 0.55$_{\pm0.05}$  & 1.51 \\ 
$^{12}$C$^{16}$O (2-0) & 21.28$_{\pm0.13}$ & 0.93 & - & $\sim$0.87 \\ 
$^{12}$C$^{16}$O (3-1) & 22.20$_{\pm0.13}$ & 0.93 & -  & $\sim$1.01 \\ 
$^{12}$C$^{16}$O (4-2) & 22.40$_{\pm0.11}$ & 0.90 & -  & $\sim$1.31 \\ 
$^{12}$C$^{16}$O (5-3) & 22.63$_{\pm0.12}$ & 0.91 & -  & $\sim$1.46 \\ 
\hline 
\\
\multicolumn{5}{c}{VX Sgr (active/quiescent phase)} \\ 
\hline\hline
\text{Region} & \text{Mean angular diameter $ \theta $ [mas]}  & \text{Scaling factor $A$} & \text{Phase shift $\Delta\phi_{\rm vis} $}  & \text{Amplitude $\Delta \theta $ [mas]}  \\  
\hline
Continuum & 9.29$_{\pm0.10}$/8.69$_{\pm0.09}$ & 0.74 & 0.56$_{\pm0.02}$/ -  & 1.17/0.36 \\
$\ion{Ti}{i}$ and $ \ion{Sc}{i}$ & 11.05$_{\pm0.32}$/11.03$_{\pm0.09}$ & 0.72 & 0.61$_{\pm0.06}$/ - & 1.03/0.60 \\ 
H$_2$O & 13.01$_{\pm0.60}$/13.50$_{\pm0.08}$ & 0.63 & 0.59$_{\pm0.05}$/ -  & 2.49/1.05 \\ 
$^{12}$C$^{16}$O (2-0) & 16.38$_{\pm0.06}$/17.01$_{\pm0.10}$ &  0.63 & - & $\sim$1.30/0.65 \\ 
$^{12}$C$^{16}$O (3-1) & 17.00$_{\pm0.06}$/17.55$_{\pm0.12}$ & 0.61 & - & $\sim$1.37/0.81 \\ 
$^{12}$C$^{16}$O (4-2) & 17.01$_{\pm0.05}$/17.57$_{\pm0.09}$ & 0.61 & - & $\sim$1.16/1.61 \\ 
$^{12}$C$^{16}$O (5-3) & 17.45$_{\pm0.06}$/17.66$_{\pm0.11}$ & 0.57 & - & $\sim$1.49/1.70 \\ 
\hline
        \end{tabular}
\end{table*}

\subsection{Angular diameters}
\label{chapter:fitting}
Due to our strategy of acquiring a large amount of time series data to constrain the variability at larger scales, each epoch includes data only in the first lobe of the visibility function (small VLTI-AT configuration) such that targets are moderately resolved at continuum ($|V|^2 > 0.2 $). It is thus not possible to use a limb-darkened model, for which we would need to cover a second lobe of the visibility function (requiring a larger VLTI-AT configuration). Instead, to fit $|V|^2$, we use a simple uniform disk (UD) geometrical model, which describes well the photospheric continuum \citep[e.g.,][]{wittkowski18}. However, the situation is more complicated for the visibility function of the molecular bands, which yield significantly larger UDs compared to the continuum and thus become partly resolved ($|V|^2 \sim 0 $). At such low $|V|^2$ near the first null point, the visibility function may show departures from UD and may be more sensitive to limb-darkening effects and surface substructures. To avoid these effects in our data, we use only values larger than $|V|^2 > 0.05$. On some dates, the molecular bands become so extended that $|V|^2 $ data points in some longer baselines lie below this limit. The situation is improved by binning our data in 2-week epochs (about $ \sim 5 \%$ of the pulsation period), which gives us more data points above this limit on some dates. 

Another major component that affects the visibility function is the contribution from the spatially over-resolved background. The origin of this contribution is disputed \citep[e.g.,][]{kluska19}. To account for this in our data, we add a scaling factor $A$ that corresponds to the over-resolved flux in the following way \citep[e.g.,][]{perrin07}:
\begin{equation}
|V|^2_{\rm model} (A, \rho, \theta_{\rm UD}  ) = A * |V|_{\rm UD}^2 (\rho,\theta_{\rm UD} ),
\end{equation}
where $|V|^2_{\rm model} (A, \rho, \theta_{\rm UD}  )$ is the visibility function of the UD model plus the spatially over-resolved background, $\rho$ is the spatial frequency $ \rho =\sqrt{ u^2 + v^2} = \frac{B}{\lambda}$, $\theta_{\rm UD}$ is the angular diameter, and $|V|^2_{\rm UD} (\rho,\theta_{\rm UD} )$ is the visibility function of a simple UD model, which is calculated from the first-order Bessel function $J_1$ \citep[e.g.,][]{born80, lawson00} as
\begin{equation}
V_{\rm UD} (\rho,\theta_{\rm UD} ) =2 \frac{J_1 ( \pi, \rho, \theta_{\rm UD} )} { \pi \rho \theta_{\rm UD}}.   
\end{equation}

\begin{figure}[t]
    \includegraphics[width=0.5\textwidth, keepaspectratio]{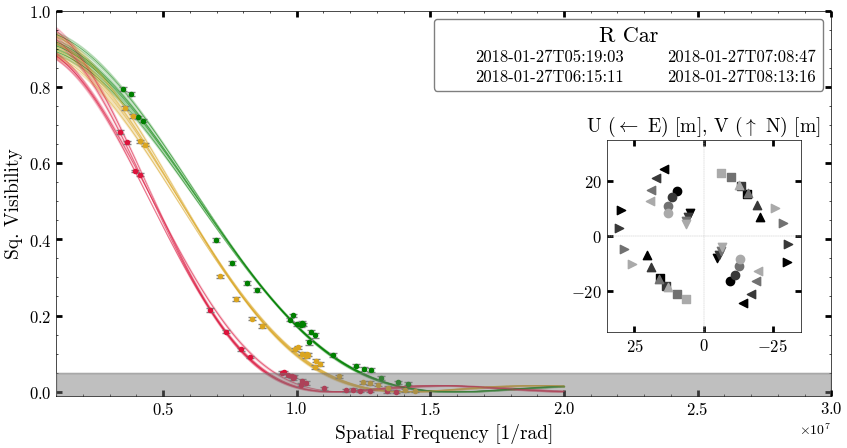}
    \includegraphics[width=0.5\textwidth, keepaspectratio]{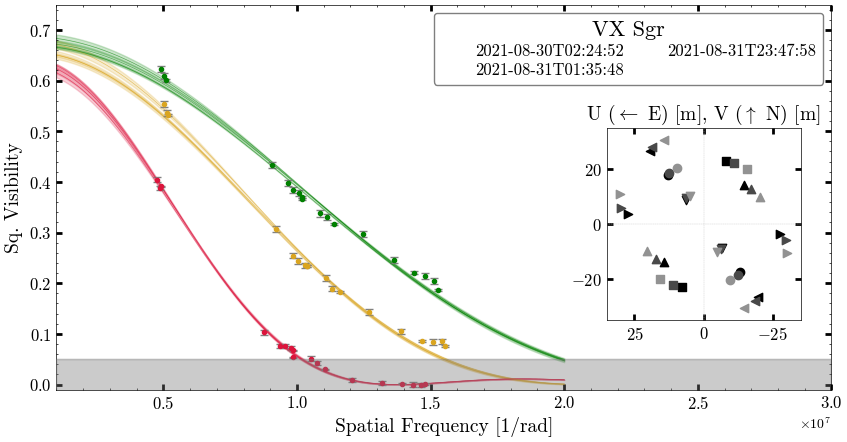}
    \caption{Examples of UD fit for observations taken within a single epoch. The fit demonstrates that the determined $\theta$ does not strongly depend on the $uv$ orientation (shown as smaller plots), and the results are mostly within error bars from each other. Different colors correspond to $ \theta_{ \rm cont} $ (green), $ \theta_{ \rm \ion{Ti}{i} , \ion{Sc}{i}} $ (yellow), and $ \theta_{ \rm CO (2-0)}$ (red). The shaded part of the graph corresponds to values below the adopted $|V|^2 = 0.05$ limit.
    {\em Upper panel:}
    R Car for four observations taken on 
    27 January 2018. See Table \ref{table:table_rcar}.
    {\em Upper panel:}
    VX Sgr for three observations taken between 
    30 and 31 August 2021. See Table \ref{table:table_vxsgr}.}
    \label{fig:ud_fit}
\end{figure}

We also explored the effect of varying $uv$ coverage on the determined $\theta$ for all considered layers, as our observations are taken at different local sidereal times, resulting in slight variations of $\rho$. Figure \ref{fig:ud_fit} demonstrates that the visibility data do not show large variations and that the determined diameters during 1-2 nights fall within error bars from each other, and thus, at our angular resolution, our targets are consistent with a UD model across different directions. 

To determine the angular diameter $\theta_{\rm UD}$ and the scaling factor $A$ at each epoch, we used the Markov chain Monte Carlo (MCMC) algorithm \citep{foreman13}. We let 250 walkers evolve in 100 steps in each wavelength region during our epochs. Before we proceed to the final fitting, we also explored the extent of contribution of the over-resolved flux in different wavelength regions. For the Mira-variable R Car, the contribution from the over-resolved flux is small at most epochs and close to unity, with $A>0.85$ (similar values for R Car in \citealt{rosales23}). Thus, we choose to set a constant scaling $A$ for the final fitting of R Car. The constant value of $A$ was calculated for each wavelength region based only on epochs with more than one observation. On the other hand, for VX Sgr, the contribution from the over-resolved component is significant and shows strong variations, with $A$ down to $\sim 0.5$ in some epochs. A similar contribution from a larger component was also determined in \citet{monnier04} and \citet{chiavassa10a}. Therefore, we fit $A$ at each epoch.

\begin{figure}[t]
    \includegraphics[width=0.5\textwidth, keepaspectratio]{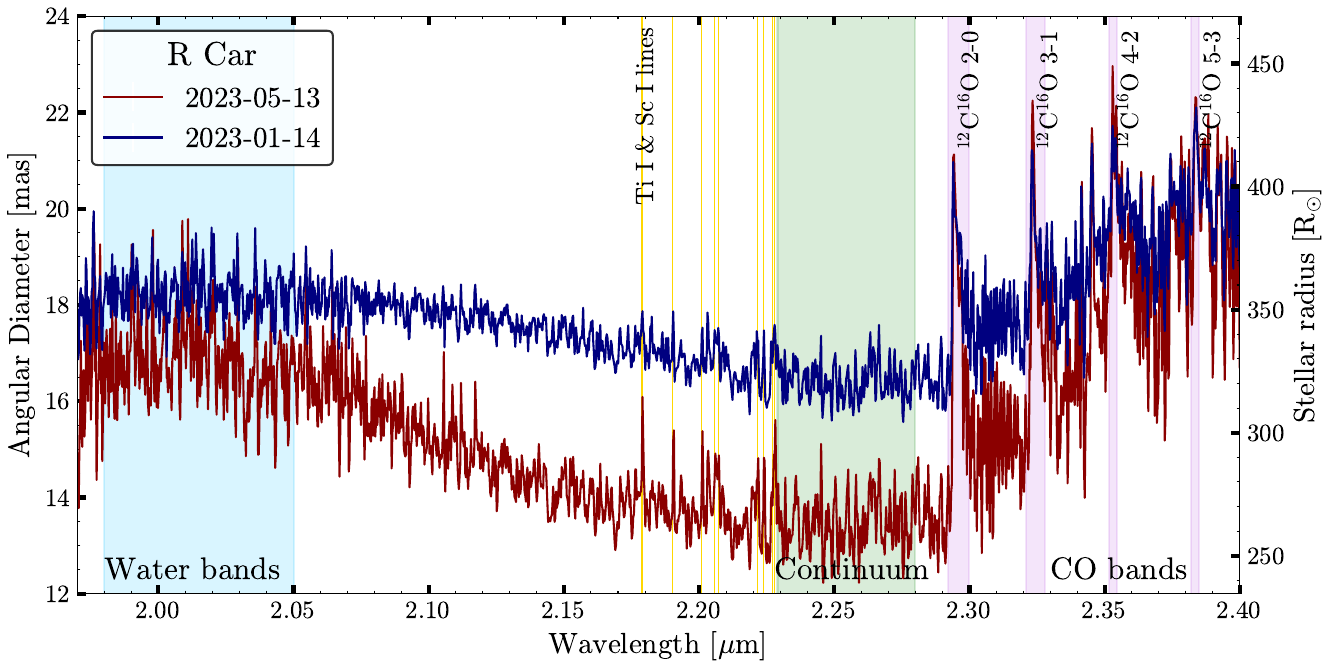}
    \includegraphics[width=0.5\textwidth, keepaspectratio]{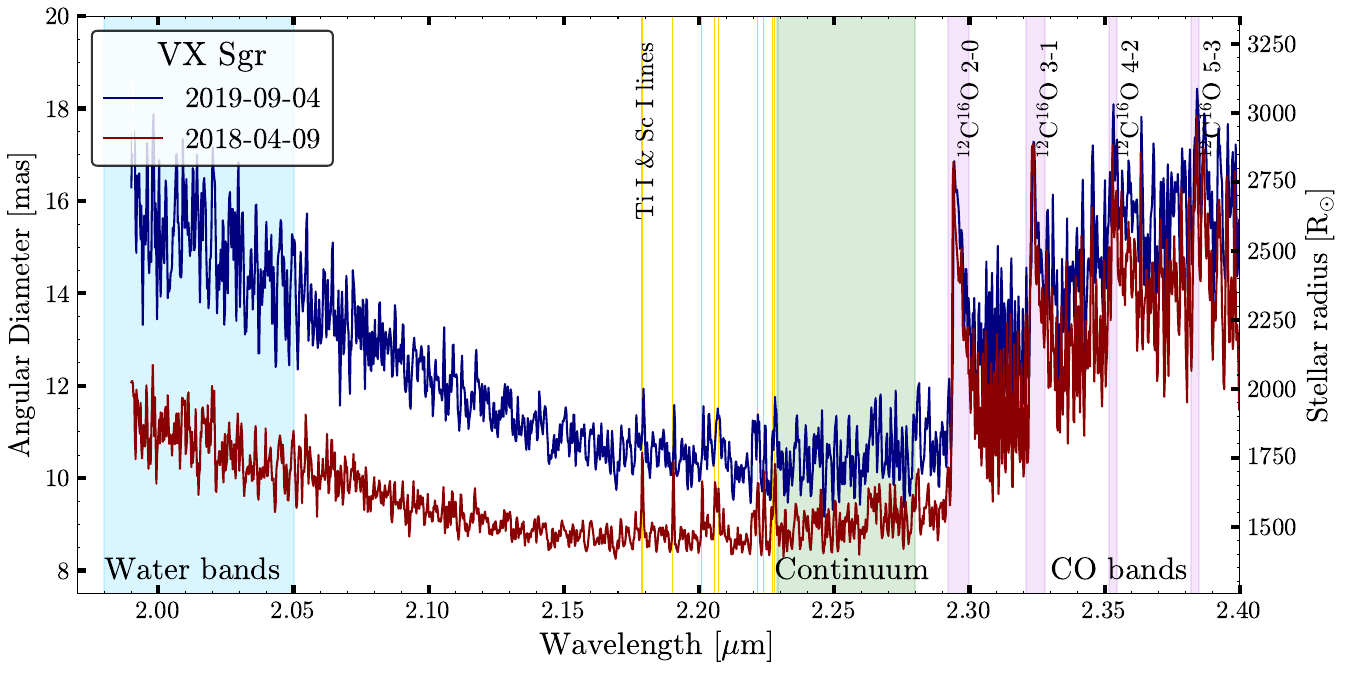}
    \caption{Comparison between the determined wavelength-dependent angular diameters during min (red) and max (blue) photospheric radius for our two targets. The observation dates are denoted in the legend. Conversion to the physical size (right y-axis) was done using the adopted distances from Table~\ref{table:table_params}.  {\em 
    Upper panel:}
    R~Car.
    {\em Lower panel:}
    VX~Sgr.
    }
    \label{fig:spectrum_ud}
\end{figure}

Table~\ref{table:table_results} lists the average value of $A$ in each wavelength region, used as a constant parameter for the final fitting of R Car and as a free parameter for VX Sgr. Determined errors for $A$ are about $\sim 0.01$. In Fig.~\ref{fig:spectrum_ud} we show an example of the determined UD diameters across the entire $K$-band wavelength range for both of our targets, at epochs corresponding to a maximum and a minimum of $\theta_{\rm cont}$.

\subsection{Variability}
\label{chapter:variability}

\begin{figure}[t]
    \includegraphics[width=0.5\textwidth, keepaspectratio]{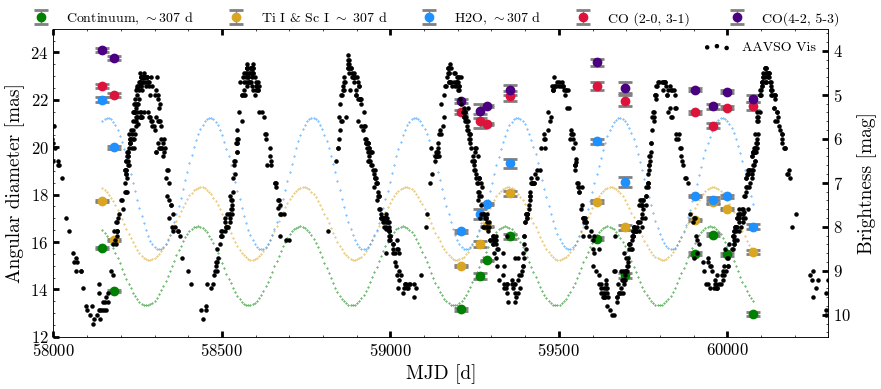}
    \includegraphics[width=0.5\textwidth, keepaspectratio]{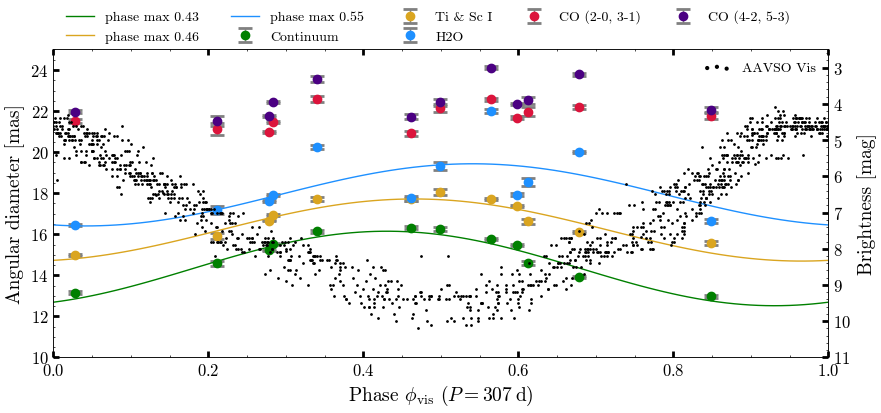}
    \caption{
    {\em Upper panel:}
    Variability of R Car based on the visual magnitude from AAVSO data and the angular diameters determined with VLTI-GRAVITY (binned in 2-week intervals). For clarity, we merged CO bands $ \theta_{ \rm CO (2-0)}$ with  $\theta_{ \rm CO (3-1)}$ and $ \theta_{ \rm CO (4-2)}$ with $\theta_{ \rm CO (5-3)}$. Different colors correspond to the continuum and to regions of the extended atmosphere.
   {\em Lower panel:}
    Phase diagram of brightness and diameter variability. Solid lines represent the sine fits for $P=307.4\rm \: d$ (same colors as atmospheric layers). Phases $ \phi_{\rm vis}$ for which the fits to angular diameters reach maximum are denoted in the caption.}

    \label{fig:series_r_car}
\end{figure}

The determined angular diameters $\theta_{\rm UD}$ and their multi-epoch variability are shown in Figs.~\ref{fig:series_r_car} and \ref{fig:series_vx_sgr}. Our results are compared to visual and $V$ light curves from AAVSO. 

Our results demonstrate the variability of continuum and molecular layers during several pulsation cycles and the relation between their variability through phase shifts. For both our targets at all epochs, the smallest angular diameters are those of continuum, and then increasing through the various atmospheric layers, i.e., $ \theta_{ \rm cont} < \theta_{ \rm \ion{Ti}{i}, \ion{Sc}{i}} < \theta_{ \rm H_2O} < \theta_{ \rm CO}$. For specific CO transitions, the situation is less clear, but at most times, layers corresponding to the highest transition also form at larger radii than the lowest transition, i.e., $ \theta_{ \rm CO (2-0)} \lesssim \theta_{ \rm CO (5-3)}$. Meanwhile, the molecular layers are more extended for the RSG VX Sgr than for the Mira R Car relative to their photospheric diameter $ \theta_{ \rm cont}$. We defined the extension of atmospheric layers with respect to the photosphere $R_{\star}$ as $\theta_{\rm layer} / \theta_{\rm cont}$ at a given epoch, assuming that $R_{\star} \simeq R_{\rm cont}$.

To estimate the mean radius of each layer, as well as the phase shift in relation to the light curve, we fitted the data with a sine curve, using the period of the light curve, similarly to Paper I. The periods are determined using the Lomb-Scargle (LS) periodogram in the \textit{astropy} package \citep{astropy}. We determined phase shifts only for such atmospheric layers that showed a similar period as the light curve (see Fig. \ref{fig:periods}). The phase shifts and mean angular diameters are listed in Table~\ref{table:table_results}. For both targets, the phase shift increases with a larger radius but becomes less regular at the most extended layers; see specific details for each target below.

\subsubsection{R Car}
For R Car, the period analysis is more straightforward, as the Mira variables are regular pulsators. There are slight variations in the amplitude and the length of each cycle (a few 
days), but this does not significantly affect our results. We thus treated R Car as a single-mode pulsator and proceeded to determine the period of pulsations. Based on the part of the AAVSO light curve covered by our VLTI observations, we derived a period of $P_{2017-2024} \sim 307.4 \: \rm d$ (see Fig. \ref{fig:periods_rcar}), while the literature gives $P_{\rm literature} = 304-314 \: \rm d$ \citep{mcdonald12, samus17, rosales23}. We used the derived period to study the phase shifts of the angular diameters of the continuum and the extended atmospheres with respect to the AAVSO light curve. 

Based on the LS periodogram in Fig. \ref{fig:periods_rcar}, we found nearly the same periods as in the light curve for the variations of the photosphere $ \theta_{ \rm cont}$ and the near-photospheric layer of $ \theta_{ \rm \ion{Ti}{i}, \ion{Sc}{i}}$ (within $1-2 \rm \: d$ difference). For the more extended molecular layer $\theta_{ \rm H_2O}$, the same period is no longer the dominant one, although it is still present in the LS periodogram, while the variations still show a relation to the light curve. As a result, it is possible to determine a phase shift. For the most extended layers of $\theta_{ \rm CO}$, similar periods as for $\theta_{ \rm H_2O}$ appear in the LS periodogram, but the variations do not follow the light curve so closely, and thus it is no longer possible to find a clear phase shift. For both $\theta_{ \rm H_2O}$ and $\theta_{ \rm CO}$, longer periods become prominent in the LS periodogram. Variations of $ \theta_{ \rm cont}$ show a maximum diameter near $\phi_{\rm vis} \sim 0.43$, i.e., roughly anticorrelated with respect to the light curve. For the neighboring layers ($ \theta_{ \rm cont} \rightarrow \theta_{ \rm \ion{Ti}{i}, \ion{Sc}{i}} $ and $ \theta_{ \rm \ion{Ti}{i}, \ion{Sc}{i}} \rightarrow \theta_{ \rm H_2O}$), the phase shift increases by $\Delta  \phi_{\rm vis} \sim 0.05$, i.e., about 2-3 weeks, with increasing distance from the photosphere. The required photospheric velocity for such a phase shift between, for example, $ \theta_{ \rm cont}$ and $\theta_{ \rm \ion{Ti}{i}, \ion{Sc}{i}}$ would be from $15$ to $20 \rm \: km s^{-1}$ (for radius difference of about $ \sim 40 \: \rm R_\odot$ from Fig.~\ref{fig:spectrum_ud}), which agrees with the expected values for Mira pulsators \citep[e.g.,][]{scholz00, kravchenko18}.  

Overall, the variations of angular diameter show a clear relation to the light curve, especially for the photosphere $ \theta_{ \rm cont}$ and near-photopsheric layers $ \theta_{ \rm \ion{Ti}{i}, \ion{Sc}{i}}$, which are roughly anticorrelated with the light curve, showing a maximum diameter close to minimum brightness. The mean extension of the outermost CO layers $\theta_{ \rm cont}$ with respect to the photosphere is $\sim 1.5 \:  R_{\star}$, but ranges between $\sim 1.3$ and $1.7 \:  R_{\star}$ during pulsation cycles. Conversion to geometrical size, using the distance and its uncertainty from Table~\ref{table:table_params}, yields $R_{\star} = 280 \pm 25 \: \rm R_\odot$ ($\sim 1.3 \: \rm au$) for the continuum, with the pulsations amplitude of $\Delta R  \sim  35 \: \rm R_\odot $. The mean extension of CO layers is about $\sim 430 \pm 38 \: \rm R_\odot$ ($\sim 2 \: \rm au$).

\subsubsection{VX Sgr}
\begin{figure}[t]
    \includegraphics[width=0.5\textwidth, keepaspectratio]{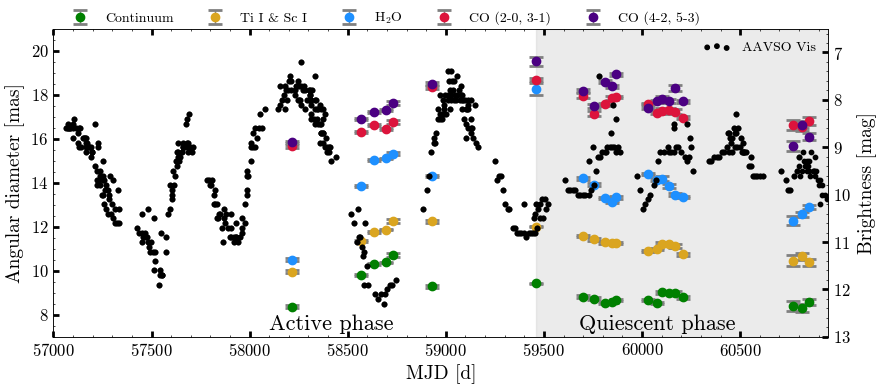}
    \includegraphics[width=0.5\textwidth, keepaspectratio]{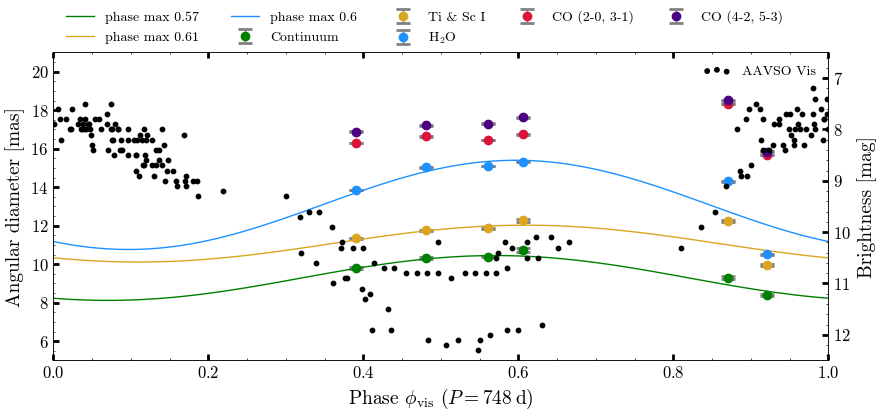}
    \caption{Same as Fig.~\ref{fig:series_r_car} but for VX Sgr. 
    {\em Upper panel:}
    Active and quiescent cycles (the latter highlighted in gray). 
   {\em Lower panel:}
    Phase diagram for VX Sgr during the active cycle. Solid lines represent the sine fits. In the phase diagrams, we did not use the extreme values of diameter from August 2021. During the quiescent cycle, it was not possible to determine the phase shift to the light curve.
    }
    \label{fig:series_vx_sgr}
\end{figure}

For RSGs, the variability is much more irregular and usually consists of several different modes of variability \citep{kiss06}. VX Sgr is an extreme RSG, perhaps it is even more irregular than other RSGs, exhibiting active and quiescent phases \citep{tabernero21}. Nonetheless, our results in Fig.~\ref{fig:series_vx_sgr} clearly show a relation of atmospheric extension with the light curve as well. First, there is an active phase with large brightness and diameter variations, ending around mid-2021 (MJD $\sim 59\,450$), followed by a quiescent phase. During the time covered by our observations, based on the light curve, we found $P_{2017-2021} \sim 748\: \rm d$ for the active phase, which agrees with the reported period of $754\pm 56 \: \rm d$ for the full light curve  by \citet{kiss06}. For the quiescent phase, we found $P_{2021-2025} \sim 316\: \rm d$ . In the following analysis, we therefore analyze pulsation properties during the active and quiescent phases separately. The cycles are highlighted in Fig.~\ref{fig:series_vx_sgr}.

\begin{figure}[htpb]
    \subfloat[]{\includegraphics[width=0.5\textwidth, keepaspectratio]{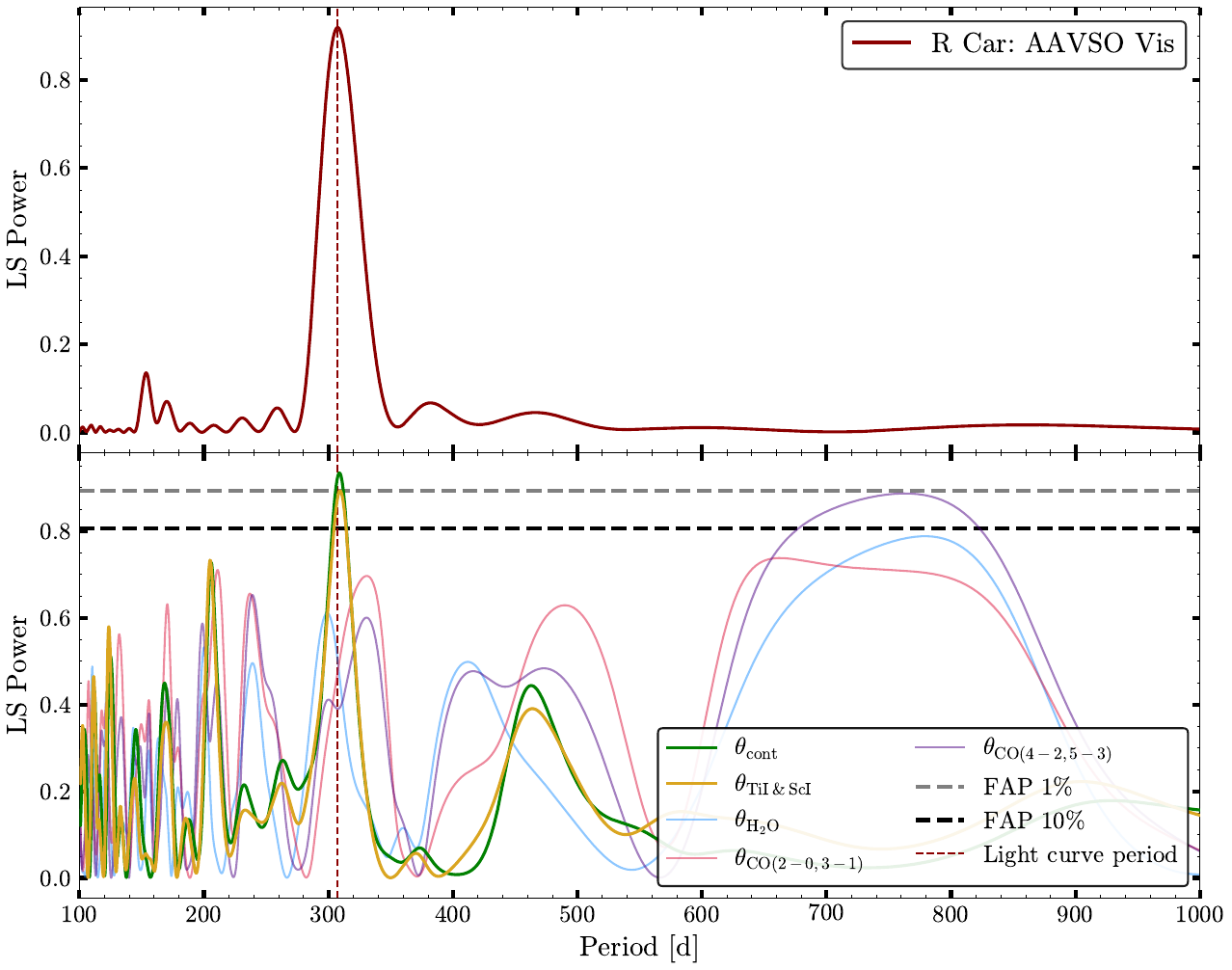}\label{fig:periods_rcar}}\\
    \subfloat[]{\includegraphics[width=0.5\textwidth, keepaspectratio]{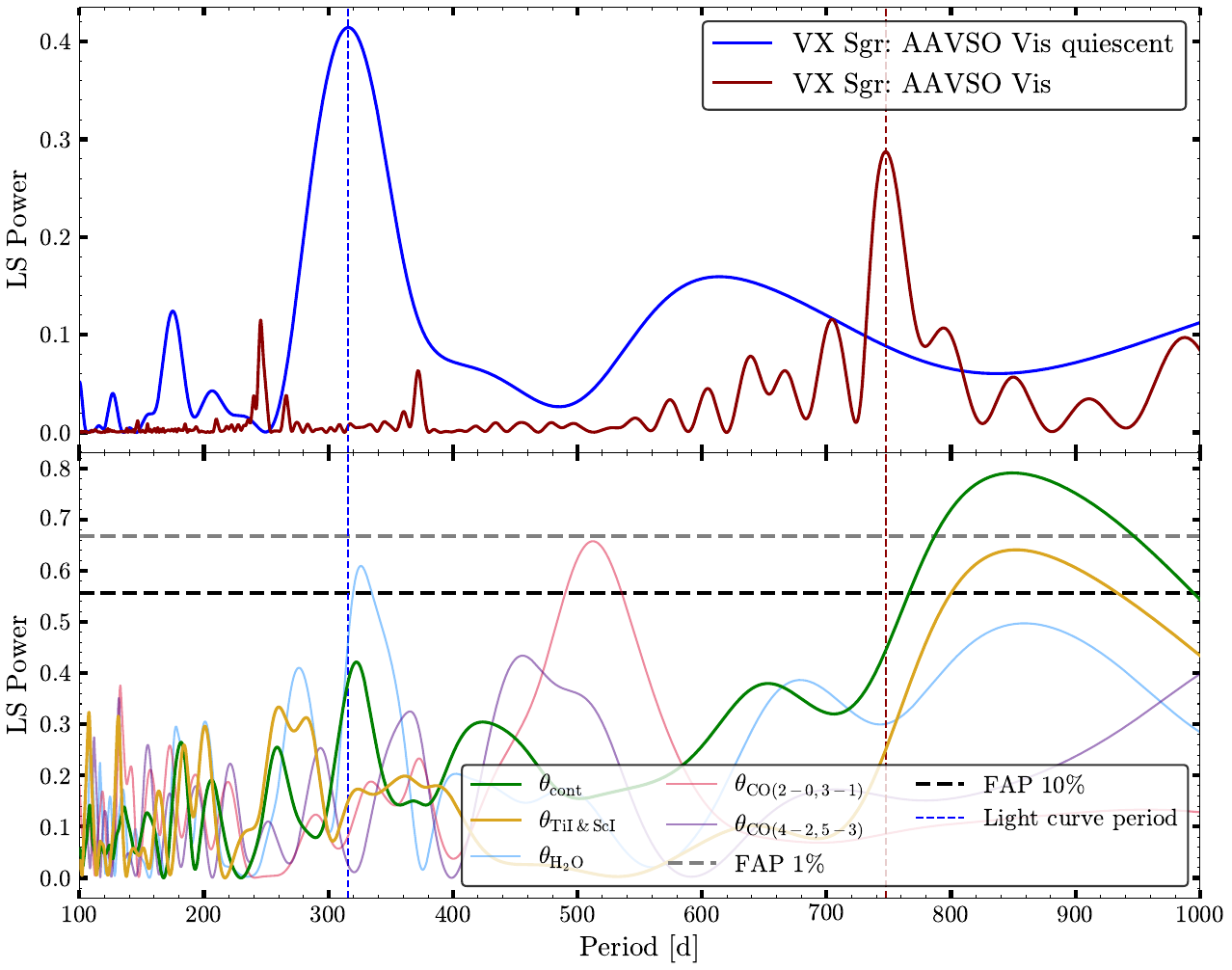}\label{fig:periods_vxsgr}}
    \caption{Lomb-Scargle normalized periodograms that we calculated for R Car (upper panels) and VX Sgr (lower panels). In the panels with diameters, we also highlight the false alarm probability \citep[FAP;][]{vanderplas18} for levels of $1\%$ and $10\%$ (dashed gray and black lines) determined with \textit{astropy}, which is one of the standard ways to judge peak significance. From our layers, usually only $\theta_{\rm cont}$ and $ \theta_{ \rm \ion{Ti}{i}, \ion{Sc}{i}}$ reach the $1\%$ threshold and show peaks similar to the AAVSO light curve. For VX Sgr, the LS periodogram is shown for the full dataset and the quiescent cycle.
    }
    \label{fig:periods}
\end{figure}

In the active cycle, $ \theta_{\rm cont}$ and $ \theta_{ \rm \ion{Ti}{i}, \ion{Sc}{i}}$ showed a similar dominant period to the light curve (see Fig. \ref{fig:periods_vxsgr}) and reach maximum diameter at $ \phi_{\rm vis} \sim 0.57-0.61$ (shift of 4-5 weeks for $ \theta_{ \rm cont} \rightarrow \theta_{ \rm \ion{Ti}{i}, \ion{Sc}{i}} $), i.e., roughly anticorrelated with respect to the light curve. Similarly to R Car, the molecular layer $\theta_{ \rm H_2O}$ showed a similar phase shift, but it has an irregular behavior and a relatively large amplitude compared to other layers, while $\theta_{ \rm CO}$ also showed a more irregular behavior. On the other hand, during the quiescent phase, variations of $ \theta_{\rm cont}$ and $ \theta_{ \rm \ion{Ti}{i}, \ion{Sc}{i}}$ showed a much smaller amplitude and did not follow the light curve so closely. Period analysis of their variations revealed period peaks relatively close to the period of the light curve of $\sim 316\: \rm d$, but not statistically significant (see Fig. \ref{fig:periods_vxsgr}). Therefore, it was not possible to determine phase shifts relative to the light curve during the quiescent phase. 

The relative atmospheric extension of the RSG VX Sgr is significantly larger compared to that of the Mira variable R Car. For VX Sgr, the extension of the outermost CO layers is typically in the range of $\sim 1.5-2.0 \: R_{\star}$, but it reached about $\sim 2.2 \:  R_{\star}$ during some cycles. The maximum extension of molecular layers was reached at the end of the active cycle, as shown by 3 observations in August 2021 (MJD$\sim 59\,450$). Surprisingly, $\theta_{ \rm H_2O}$ reached a similar angular diameter as that of $\theta_{ \rm CO}$, $\sim 2.2 \:  R_{\star}$, which is about twice compared to the value of $\theta_{ \rm H_2O}$ in 2018, where it instead showed a very weak extension. Meanwhile, since 2020, $ \theta_{\rm cont}$ and $ \theta_{ \rm \ion{Ti}{i}, \ion{Sc}{i}}$ have contracted compared to their diameters in the active phase, while showing much smaller diameter variations. Based on the distance value from Table~\ref{table:table_params}, we estimate the physical size of VX Sgr in different cycles. During the active cycle, the mean continuum radius was $R_{\star} = 1556 \pm 110 \: \rm R_\odot$ ($\sim 7 \: \rm au$), with pulsation amplitude $\Delta R  \sim  197 \: \rm R_\odot $. During the quiescent phase, the star contracted, showing the mean radius of $R_{\star}= 1456 \pm 108 \: \rm R_\odot$ and a smaller $\Delta R$ of $ \sim  60 \: \rm R_\odot $. This demonstrates extreme change to the pulsation properties of the star between the active and quiescent cycles. The mean radius of CO extension was about $2850 \pm 200 \: \rm R_\odot$ ($\sim 13 \: \rm au$). Moreover, during the extreme expansion, $R_\star$ reached a maximum of $1798 \pm 127 \: \rm R_\odot$ in September 2019, followed by a maximum of H$_2$O and CO layers up to about $\sim 3200  \: \rm R_\odot $ observed in August 2021. This extreme event is further discussed in Sect. \ref{chapter:discussion}.

\section{Comparison to the CO5BOLD simulation}
\label{chapter:models}

\subsection{Extracting interferometric observables}
We compared our observations to the CO5BOLD simulations \citep{freytag12}, which, along with Athena++ \citep{stone20, goldberg22}, are one of the few 3D RHD models of AGB and RSG stars available in the community \citep{chiavassa24}. A CO5BOLD model is the most suitable for comparison to our results, similar to several previous works, e.g., \citetalias{kravchenko20} and \citet{chiavassa22}. The computation of several pulsation cycles of such 3D models is computationally expensive; thus, the grid is not dense enough to provide a model with exactly the same parameters as R Car (see Table~\ref{table:table_params}). From the extended grid of 3D models of low-mass AGB stars used for pulsation studies \citep{ahmad25}, all relying on the gray opacity approximation, we selected the st28gm05n057 model, which is close to the physical parameters of R Car, except for longer $P$. Namely, $M = 1 \: \rm M_{\odot}$, $R = 291 \: \rm R_{\odot}$, $L = 5024  \: \rm L_{\odot}$, $T_{\rm eff} = 2847  \: \rm K$, $\log(g) = -0.493 \: \rm (cgs)$, and $P = 375.1  \: \rm d$. This newer model was also chosen as the most suitable because it shows a similar behavior in terms of Mira-like regular pulsations, while the majority of CO5BOLD models show more irregular behavior. The same is the case for the few available RSG models, which generally do not reproduce the interferometric observations so well \citep[e.g.,][]{arroyo15} and are usually calculated for smaller masses than RSGs have. Therefore, in this work, we focus only on comparison for the selected AGB model, which shows the best agreement with the observed variability.

\begin{figure*}[t]

    \includegraphics[width=0.44\textwidth, keepaspectratio]{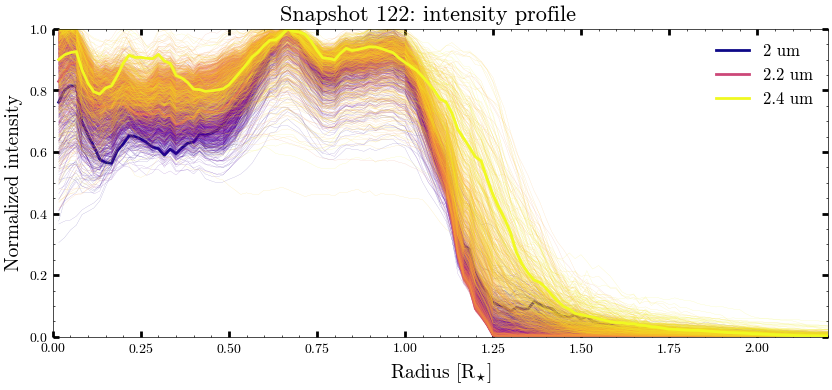}
    \includegraphics[width=0.55\textwidth, keepaspectratio]{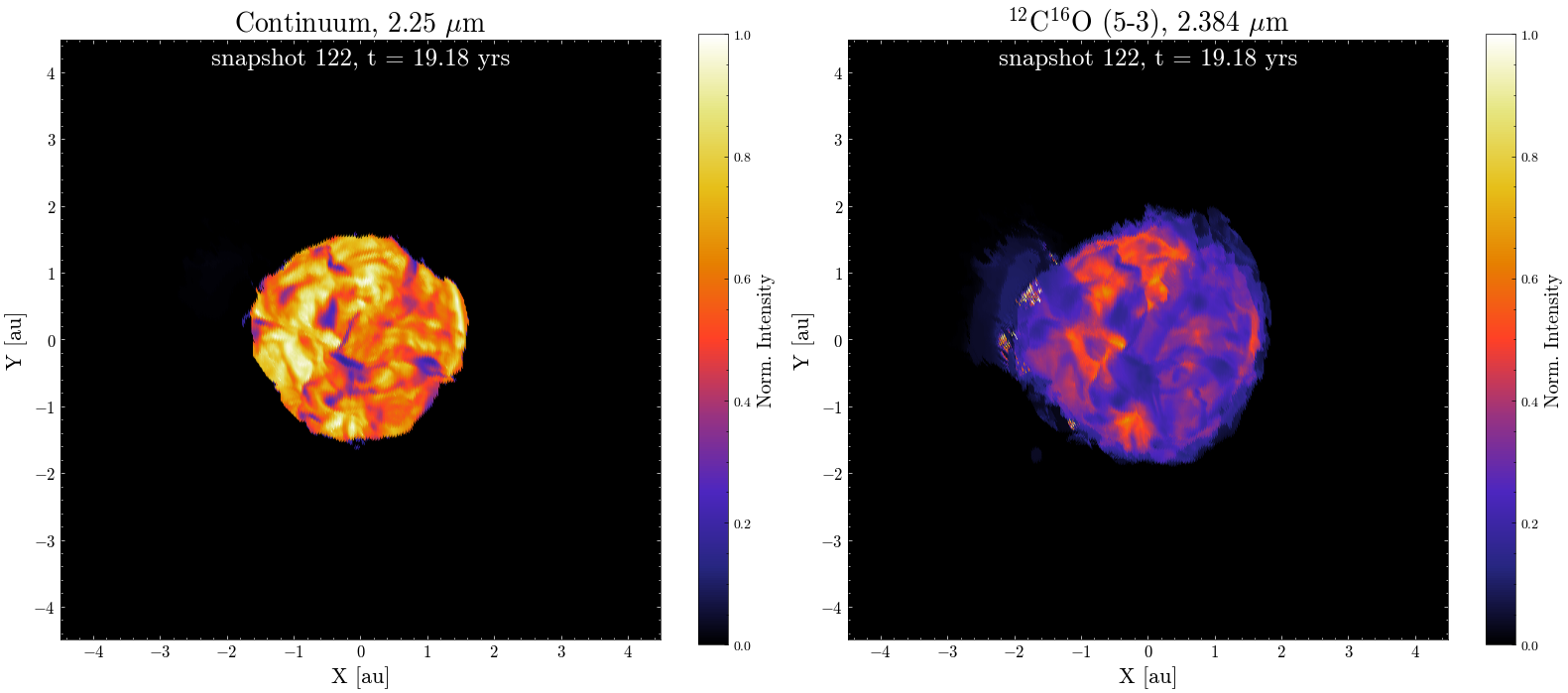}
    
    \includegraphics[width=0.44\textwidth, keepaspectratio]{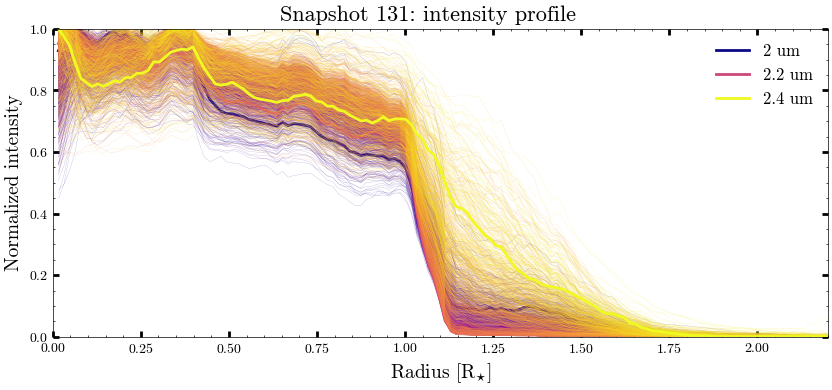}
    \includegraphics[width=0.55\textwidth, keepaspectratio]{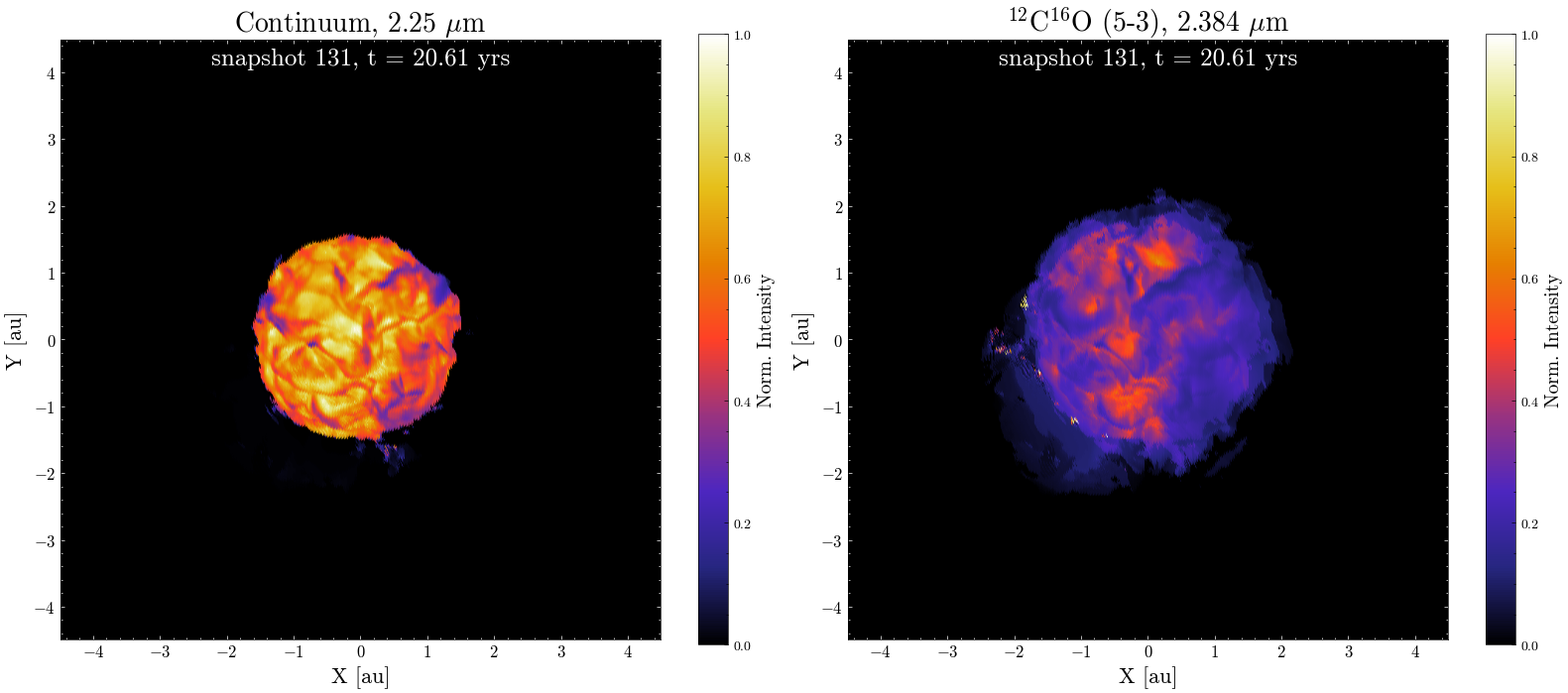}

    \caption{Intensity profiles and intensity maps for snapshots 122 ($t=19.18 \:  \rm yrs$) and 131 ($t=20.61 \:  \rm yrs$) of the AGB model st28gm05n057.
    {\em Left panels:}
    Intensity profiles plotted for all wavelengths in the GRAVITY spectral range. Three representative wavelengths are highlighted.
    {\em Right panels:}
    Intensity maps for two selected representative wavelengths, $2.250 \: \rm \mu m $ (continuum) and $2.384 \: \rm \mu m $ (band head of $^{12}$C$^{16}$O, 5-3).  A linear scale was used for the intensity. The full movie is available in electronic format at the following link: \url{https://doi.org/10.1051/0004-6361/202659237}.}
    \label{fig:3d_models}
\end{figure*}

For the selected model st28gm05n057, we identified a part of the model light curve that resembled the observed photometric variability the most. 
Then, for the selected snapshots, we calculated the intensity in $K$-band (at $R \sim 4000$, which is equal to the spectral resolution of VLTI-GRAVITY) in a similar way as described in \citet{chiavassa09}, see Fig.~\ref{fig:3d_models}. To determine intensity profiles, we constructed rings around the center of the star, at regularly spaced distances in $ r/{R_{  \rm Ross}}$, where $ {R_{ \rm Ross}}$ is the Rosseland radius of the star. Then, we calculate $ \mu = \cos{\varphi} $, where $\varphi$ is the angle between the radial direction and the line of sight, and is related to the radius as $ r/R_{ \rm Ross} = \sqrt{1 - \mu^2 } $. The observed $|V|^2$ in the first lobe have been shown to be consistent with a UD fit for both our targets (see Fig. \ref{fig:ud_fit}), so we approximate the simulations in the same way. Most importantly, our aim was to study the overall pulsation variations and not the asymmetry of the model. Therefore, following a similar approach as in, e.g., \citet{davies00} and \citet{wittkowski01}, we use the Hankel transform. We calculated the interferometric visibility of the model,  based on $I ({\mu})$, angular diameter $\theta$, and spatial frequencies $B/\lambda$ corresponding to our observations, i.e., for baselines from 5 to 35 m.

Then, we apply the same fitting procedure as for the observations in Sect. \ref{chapter:fitting} to obtain the angular diameters of continuum, as well as atomic and molecular species. We also determined periods and phase shifts.

\begin{figure}[t]
    \includegraphics[width=0.5\textwidth, keepaspectratio]{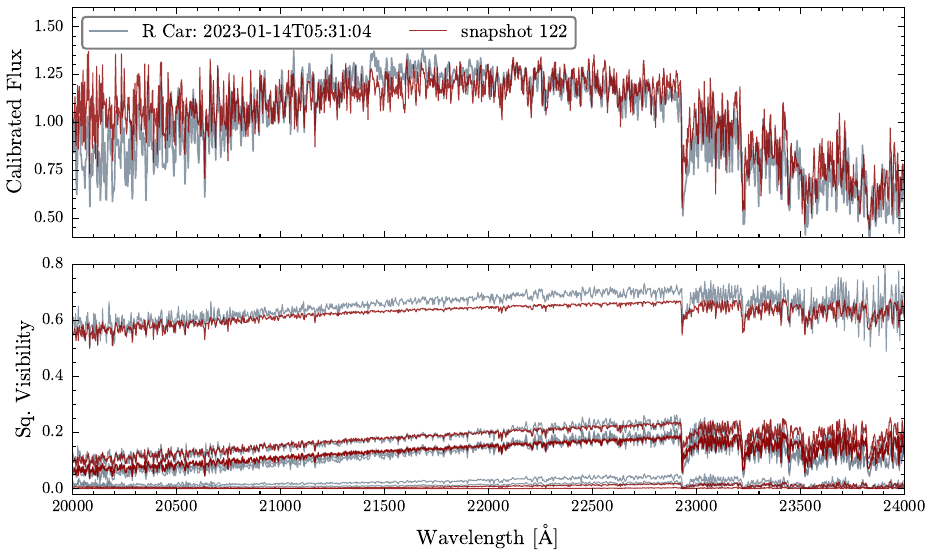}
    \caption{Example of a snapshot of the AGB model that fits the observations of R Car well. The observation of R Car is the same as shown in Fig. \ref{fig:spectrum}, i.e., 
    14 January 2023, and it is fitted with snapshot 122 for $\theta = 14.87 \: \rm mas$. 
    {\em Upper panel:}
    Calibrated flux of R Car and the flux of the AGB model.
    {\em Lower panel:}
    Squared visibility amplitude $|V|^2$ of R Car. Six baselines are shown in the same color and compared with the synthetic $|V|^2$ of the AGB model for the same baselines.
    }
    \label{fig:model_fit}
\end{figure}

\begin{figure}[t]
    \includegraphics[width=0.5\textwidth, keepaspectratio]{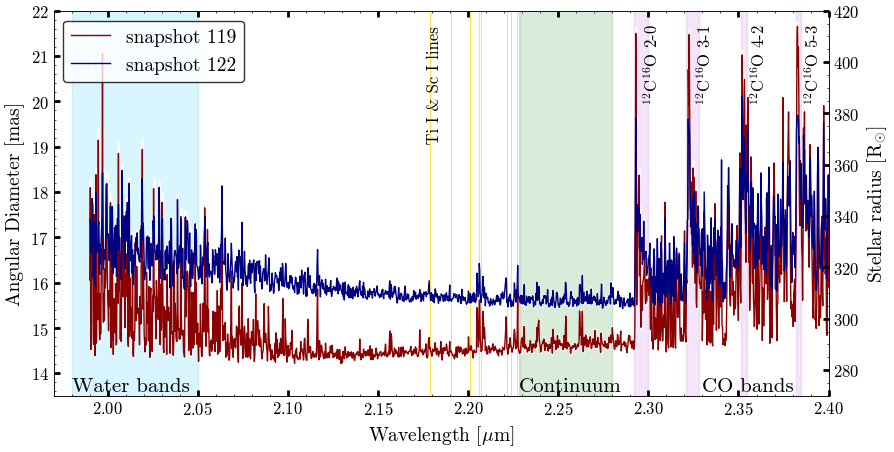}
    \caption{Same as Fig. \ref{fig:spectrum_ud} but for the synthetic angular diameters computed from snapshots 119 and 122, i.e., for the minimum (red) and maximum (blue) photospheric radius of our AGB model. Compared to the observations (Fig.~\ref{fig:spectrum_ud}), the H$_2$O bands show a weaker extension.}
    \label{fig:spectrum_ud_3d}
\end{figure}

\subsection{AGB model}

\begin{figure}[t]
    \includegraphics[width=0.5\textwidth, keepaspectratio]{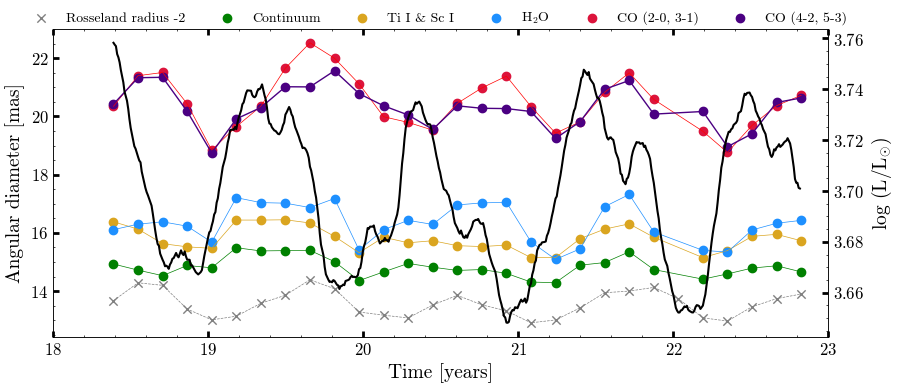}
    \includegraphics[width=0.5\textwidth, keepaspectratio]{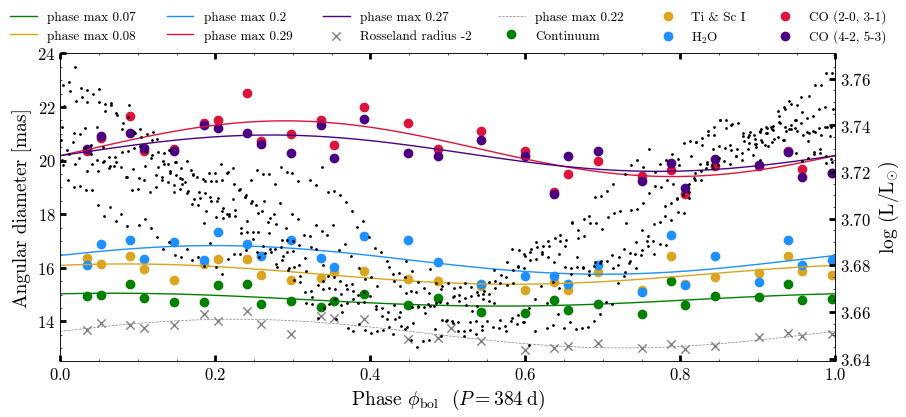}
    \caption{
    {\em Upper panel:}
    Selected part of the synthetic bolometric light curve (black line, smoothed), including the computed angular diameters for 28 snapshots of our AGB model. The base value of the angular diameter was set to be similar to that of R Car. The Rosseland radius was converted to the angular diameter using the distance of R Car (vertically shifted by -2 mas).
   {\em Lower panel:}
    Phase diagram showing the phase shifts between the synthetic light curve (black dots) and the angular diameters for different atmospheric layers.}

    \label{fig:3d_models_series}
\end{figure}

For the AGB model st28gm05n057, we extracted 28 snapshots (snapshot numbers from \_117 to \_145), covering $\sim 18-23 \: \rm yr$, i.e., 4 pulsation cycles. Example of the calculated observables and their comparison to observations for one of the selected snapshots are shown in Fig. \ref{fig:model_fit}. The comparison shows a close agreement between the computed flux and $|V|^2$ for the continuum region and the CO bands. The resulting atmospheric extension of the model in the full $K$-band is shown in Fig.~\ref{fig:spectrum_ud_3d}. The variability of the model st28gm05n057 is shown in Fig.~\ref{fig:3d_models_series}, and the results are listed in Table~\ref{table:table_models_results}.

We determined the period of the synthetic bolometric light curve during the cycles covered by our snapshots, and we found $P=383.5 \; \rm d$. This allowed us to determine phase shift with respect to the bolometric light curve, $\phi_{\rm bol}$. We note that $\phi_{\rm bol} \approx \phi_{\rm vis}$, as also shown in previous studies \citep[e.g.,][]{kravchenko19}. We then calculated a phase diagram, similar to that for our observations of R Car, and compared the phase shifts $\phi_{\rm bol}$. First, we compared the variability of $R_{ \rm Ross}$ to the bolometric light curve, yielding a maximum diameter at $\phi_{\rm bol} \sim0.22$, preceding the minimum of the light curve. The variations of synthetic angular diameter also follow the light curve, although, for variations of $ \theta_{ \rm cont}$ and $ \theta_{ \rm \ion{Ti}{i}, \ion{Sc}{i}}$, we found even smaller phase shifts, with maximum diameters approximately at $\phi_{\rm bol} \sim 0.07-0.08$. The phase shift between the photosphere and atomic layers ($ \theta_{ \rm cont} \rightarrow \theta_{ \rm \ion{Ti}{i}, \ion{Sc}{i}} $) is smaller than for R Car, i.e., only $\Delta\phi_{\rm bol} \sim 0.01$. Nonetheless, we note that during the last two pulsation cycles, $ \theta_{\rm cont}$ shows a better agreement with variations of $R_{ \rm Ross}$, shifting the maximum of $ \theta_{\rm cont}$ to $\phi_{\rm bol} \sim 0.12$. The outer layers of H$_{2}$O and CO exhibit a regular behavior, allowing us to determine phase shifts. Molecular bands of H$_{2}$O are much weaker compared to our observations, while $ \theta_{\rm H_{2}O}$ appear to show similar variations as $ \theta_{ \rm cont}$ and $ \theta_{ \rm \ion{Ti}{i}, \ion{Sc}{i}}$, with a maximum at $\phi_{\rm bol} \sim 0.2$. Unlike for the observations, molecular bands of CO also showed a clear relation to the light curve, with maximum diameters at $\phi_{\rm bol} \sim 0.3$ for variations of $ \theta_{\rm CO}$. 

Overall, for the Mira AGB model st28gm05n057, variations of synthetic angular diameters show a clear relation to the light curve, although the pulsation amplitudes are much smaller than for the observations (see Tables \ref{table:table_results} and \ref{table:table_models_results}). The model shows a smaller relative atmospheric extension compared to R Car, especially at H$_{2}$O bands, which show extension only up to $\sim 1.1 \: R_{\star}$. The extension of the outermost CO layers of the model with respect to the photosphere is better reproduced, showing a mean extension of $\sim 1.4 \: R_{\star}$, but it ranges between $\sim 1.3$ and $1.5 \:  R_{\star}$ during pulsation cycles and shows a more regular behavior. Full intensity maps in Fig.~\ref{fig:3d_models} give a more complete picture. The star clearly shows a larger radius at the wavelengths corresponding to CO band heads (as also shown in Fig.~\ref{fig:spectrum_ud_3d}), while during each pulsation cycle, non-spherical outflows occur.

\begin{table*}[t]
        \centering
        \caption{Results related to the angular diameters of the AGB model and phase shifts (with respect to $\phi_{\rm bol}$) determined in the same way as for the observations.} 
        \label{table:table_models_results} 
        \setlength{\extrarowheight}{1pt}
        \begin{tabular}{ccccc}

\hline \hline
\text{Region} & \text{Mean angular diameter $ \theta $ [mas]}  & \text{Phase shift $\Delta \phi_{\rm bol}  $}   & \text{Amplitude $\Delta \theta $ [mas]}  \\ 
\hline
Rosseland radius & -  & 0.22$_{\pm0.02}$  & - \\ 
Continuum & 14.81$_{\pm0.06}$  & 0.07$_{\pm0.05}$  & 0.24 \\ 
$\ion{Ti}{i}$ and $ \ion{Sc}{i}$ & 15.76$_{\pm0.06}$  & 0.08$_{\pm0.04}$ & 0.38 \\
H$_2$O & 16.29$_{\pm0.10}$  & 0.20$_{\pm0.04}$ & 0.53 \\ 
$^{12}$C$^{16}$O (2-0) & 20.35$_{\pm0.09}$  & 0.27$_{\pm0.02}$ & 0.95 \\ 
$^{12}$C$^{16}$O (3-1) & 20.53$_{\pm0.16}$  & 0.30$_{\pm0.03}$ & 1.14 \\
$^{12}$C$^{16}$O (4-2) & 19.99$_{\pm0.10}$  & 0.22$_{\pm0.03}$  & 0.74 \\ 
$^{12}$C$^{16}$O (5-3) & 20.56$_{\pm0.12}$ &  0.32$_{\pm0.04}$  & 0.68   \\

\hline 
        \end{tabular}
\end{table*}

\section{Discussion}
\label{chapter:discussion}

\subsection{Comparison to other works}
Our results are compatible with other works on the atmospheric extension of AGBs and RSGs, with reported maximum CO extensions up to $\sim 1.5-2 \: R_{\star}$ in near-IR, obtained with AMBER/GRAVITY, \citep[e.g.,][]{arroyo15, wittkowski18}. Meanwhile, in mid-IR bands covered by MATISSE, there are more extended layers of H$_2$O, CO, and SiO, up to $\sim 10 \:  R_{\star}$ \citep[e.g.,][]{chiavassa22, gonzalez24}. As recently shown by \citet{gonzalez23, gonzalez24} using GRAVITY/MATISSE data, for a star of given stellar parameters, higher mass-loss rates are correlated with larger atmospheric contribution from molecular bands, altering the density profile in outer layers.

Regarding the variability and phase shifts, for the Mira R Peg in Paper I, it was found that $\theta_{\rm cont}$ showed a maximum at $\phi_{\rm vis} \sim 0.57$, while more extended layers showed longer phase shifts with respect to the light curve. Meanwhile, using a longer time coverage in this paper, for the Mira R Car, we found $\phi_{\rm vis} \sim 0.43$ for the maximum of $\theta_{\rm cont}$. We note that both these results are consistent with published high-cadence observations over several pulsation cycles for the Mira S Lac using PTI \citep{aguirre24}, where diameter variations revealed roughly a plateau for the maximum $\theta_{ \rm cont}$ between $\phi_{\rm vis} \sim 0.3$ and $0.6$, while they also reported a regular behavior for the spectral regions dominated by $\rm H_2O$ and $\rm CO$. However, the spectral resolution of PTI is much lower than that of GRAVITY, with only 5 spectral channels in $K$-band, i.e., a width of $\sim 0.08 \: \rm \mu m$ per channel. Therefore, their results for the $\rm CO$ layer are not fully comparable, as in our work, we measured specific band heads of $\rm CO$. We also note that a sine fit may not be a precise representation of the pulsations of radius and light curve; thus, the determined phase shifts are approximations. In any case, for R Car, the regular variations of the photospheric diameter do not show large deviations from the fit. For RSGs, there are no models that can fully reproduce the pulsation curves. However, recently, first attempts have been made to model radial envelope pulsations of pre-SN RSGs \citep{suzuki25, bronner25, laplace26, farag26}. 

Additionally, we compared our observations to synthetic $K$-band data calculated for our AGB CO5BOLD model st28gm05n057  (see Sect. \ref{chapter:models}). For the photospheric radius (defined as Rosseland radius of the model), we found a maximum at $\phi_{\rm bol} \sim 0.22$, and at even earlier $\phi_{\rm bol}$ when using interferometric observables. We found a similar value for the AGB model st29gm06n001, used in \citetalias{kravchenko20}, which showed $\phi_{\rm bol} \sim 0.2-0.3$, while the RSG model st35gm04n38, used in \citet{kravchenko19}, showed $\phi_{\rm bol} \sim 0.3-0.4$. Additionally, in previous works using the CODEX models \citep[e.g.,][]{ireland11}, maximum photospheric diameter was also usually reached for $\phi_{\rm bol}$ between 0.3 and 0.4. Therefore, all these previous modeling works suggest a maximum photospheric radius preceding the minimum brightness. In our AGB model, the outer layers of H$_{2}$O and CO exhibit a more regular behavior than the observations of R Car and VX Sgr (see Sect. \ref{chapter:variability}). The new generation of CO5BOLD models also show a similar irregular behavior of more extended layers up to $\sim 2 \: R_{\star}$ (velocity or density plots vs radius and time), i.e., for AGB models in \citet{freytag23} and for RSG models in \citet{freytag24}.

We also investigated the atmospheric extension of molecular species in our AGB model. We found a good agreement with the CO bands, as the AGB model showed comparable extension to observations in all snapshots. However, there is a significant discrepancy in the H$_2$O bands, both in the flux and the visibility. The contribution from H$_2$O bands appears much weaker in the models compared to our observations, especially when comparing Figs.~\ref{fig:spectrum_ud} and \ref{fig:spectrum_ud_3d}. The slope of the spectrum and visibility below $ \sim 2.1 \: \rm \mu m $ and above $ \sim 2.3 \: \rm \mu m $, where the weaker bands of H$_2$O should contribute, does not correspond to the observations. This difference could be due to the gray Rosseland opacity approximation in the CO5BOLD model, which may result in a higher temperature in the upper atmosphere and thus prevent the formation of some molecular species and dust \citep{hofner19}. Indeed, \citet{freytag23} published a few models with an extended computational box and with wavelength-dependent opacity to properly reproduce molecular formation and dust-driven wind. Such models would be best suited for future comparison to observations, once the model grid becomes more compatible with the actual observable stars. The model grid we used in this work was constructed for pulsation studies and shows reasonable agreement with observed pulsation periods and mode properties \citep{ahmad23,ahmad25}. 

\subsection{Extreme mass-loss event of VX Sgr}
The RSG VX Sgr exhibits different pulsation properties during its active and quiescent cycles (see Sect. \ref{chapter:variability}). The determined diameters showed an extreme expansion of $\theta_{ \rm H_2O}$ and $\theta_{ \rm CO}$ at the end of the active cycle in mid-2021, the largest such expansion in our data. 
This shows some similarity to the Great Dimming event of Betelgeuse \citep{dupree22, macleod23}, where a similar behavior was observed for the photospheric layers using the tomographic method \citep{kravchenko21, jadlovsky24}, while following the event, the mode of pulsations changed from the fundamental mode to the first overtone. 

To further understand this event, we analyzed the optical spectra from the STELLA telescope. The radial velocity was determined using the cross-correlation function (CCF), in a similar way as described in \citet{jadlovsky24}. The radial velocity and full width at half maximum (FWHM) of CCF (approximately, this is the average width of spectral lines in the optical spectra) are both compatible with the scenario of a mass-loss event induced by a shock (see Fig.~\ref{fig:vx_sgr_rv}). Similar to the case of Betelgeuse \citep{kravchenko21, jadlovsky24}, two extreme shocks (as indicated by an increase in FWHM of $\sim 30 \; \rm km s^{-1}$) preceded the major outflow event. Surprisingly, within the time window from mid-2020 to mid-2021, a very strong Balmer emission emerged (see Fig.~\ref{fig:vx_sgr_rv}). Simultaneously, we also discovered that in the March 2020 observations, hydrogen Br$\gamma$ uniquely appeared and temporarily became very prominent in the spectro-interferometric data, namely in the differential phase but not in $|V|^2$ or closure phase.

This Br$\gamma$ hydrogen line is usually a signature of accretion, such as in young accreting objects \citep[e.g.,][]{gravity23, gravity24}, and it is also used to study massive interacting binaries \citep[e.g.,][]{klement25}. It has not been detected for single RSGs. Only recently it has been detected for VV Cephei type RSG binary KQ Pup by \citet{jadlovsky25}, which is the first such detection for a RSG with VLTI-GRAVITY. It was also detected in the spectrum of {the extreme RSG WOH G64 \citep{munoz24}, which is likely also a binary. In these systems, Br$\gamma$ traces the accreting hot companions and/or their circumstellar disk. However, upon further analysis of the behavior of this mysterious emission during 2020-2021, it appears that in the VX Sgr system, this may not be the same scenario. The radial velocity of the emission peak appears to follow the same curve as the radial velocity of VX Sgr (see Fig.~\ref{fig:vx_sgr_rv}), which suggests this event originates in the atmosphere of VX Sgr and does not trace a binary motion. We note that \citet{tabernero21} also reported on occasional Balmer emission, such as in 2018, but our data show much stronger Balmer emissions, while Br$\gamma$ is not present in the VLTI data from 2018. 

As already shown for R Car (Fig.~\ref{fig:spectrum_co}), hydrogen emission lines can appear as part of their pulsation cycle, due to the passing of a strong shock \citep[e.g.,][]{fox84, fabas11}. This scenario seems much more likely, as we have also detected signatures of strong shocks in the optical spectra. Thus, we conclude that indeed, the 2020-2021 event was likely caused by an extreme mass-loss event and not by a companion. We note that even though VX Sgr was hypothesized to be a super-AGB star \citep{tabernero21}, it is very unlikely, as based on its high luminosity, it should rather be an extreme RSG within the mass range of $20-40 \: \rm M_\odot$ \citep{arroyo13, arroyo15}. Indeed, we compared the published RSG pulsation models \citep{joyce20} to our determined radius of VX Sgr ($R_{\star, \rm active} = 1556 \pm 110 \: \rm R_\odot$ and $R_{\star, \rm quiescent}= 1456 \pm 108 \: \rm R_\odot$), as well as to the ratio of likely fundamental mode (FM, $\sim 748 \: \rm d$) and first overtone (O1, $\sim 316 \: \rm d$) of $\rm O1 /FM \sim 0.42$. It is clear that only the most massive RSG models would reproduce such properties.

\begin{figure}[t]
    \includegraphics[width=0.5\textwidth, keepaspectratio]{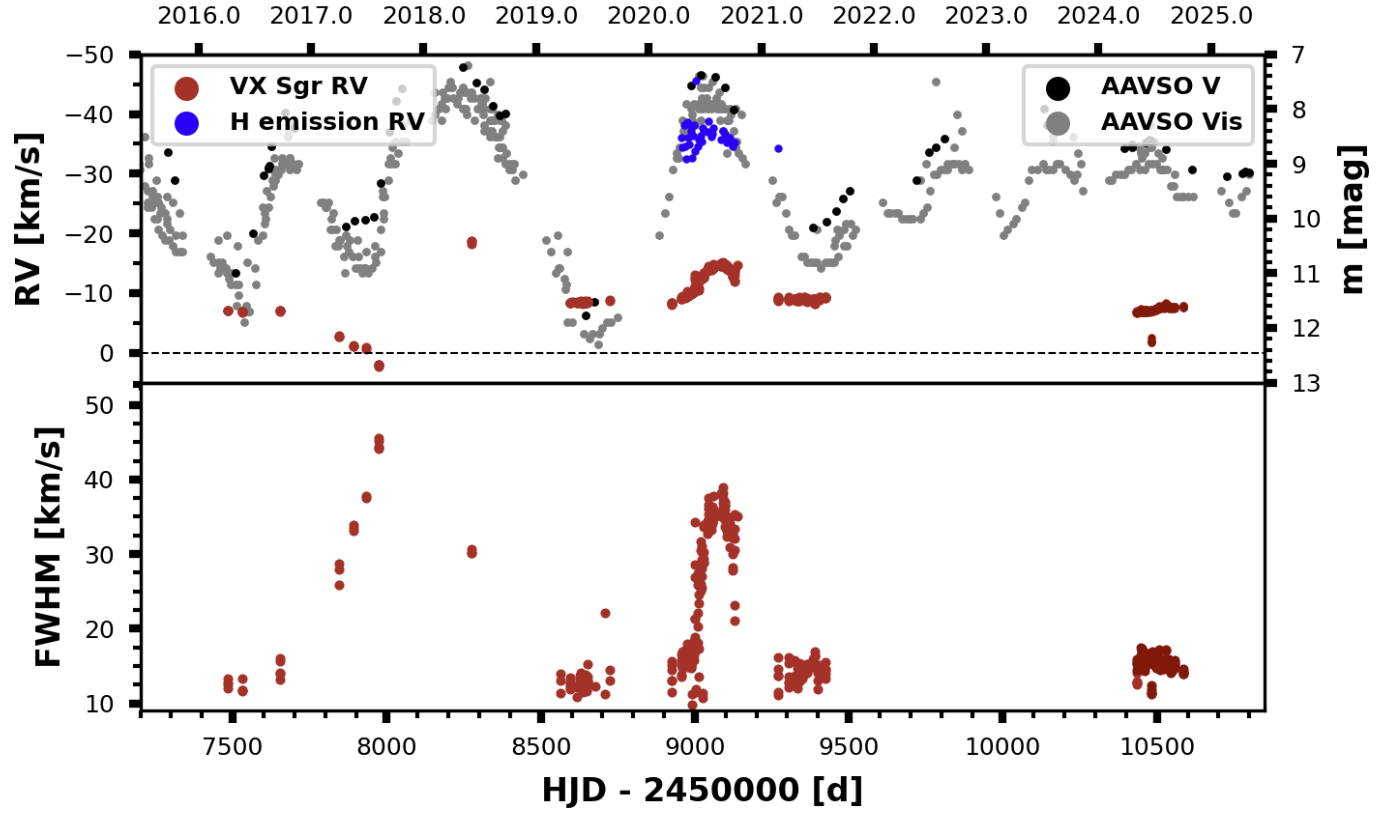}
    \includegraphics[width=0.47\textwidth, keepaspectratio]{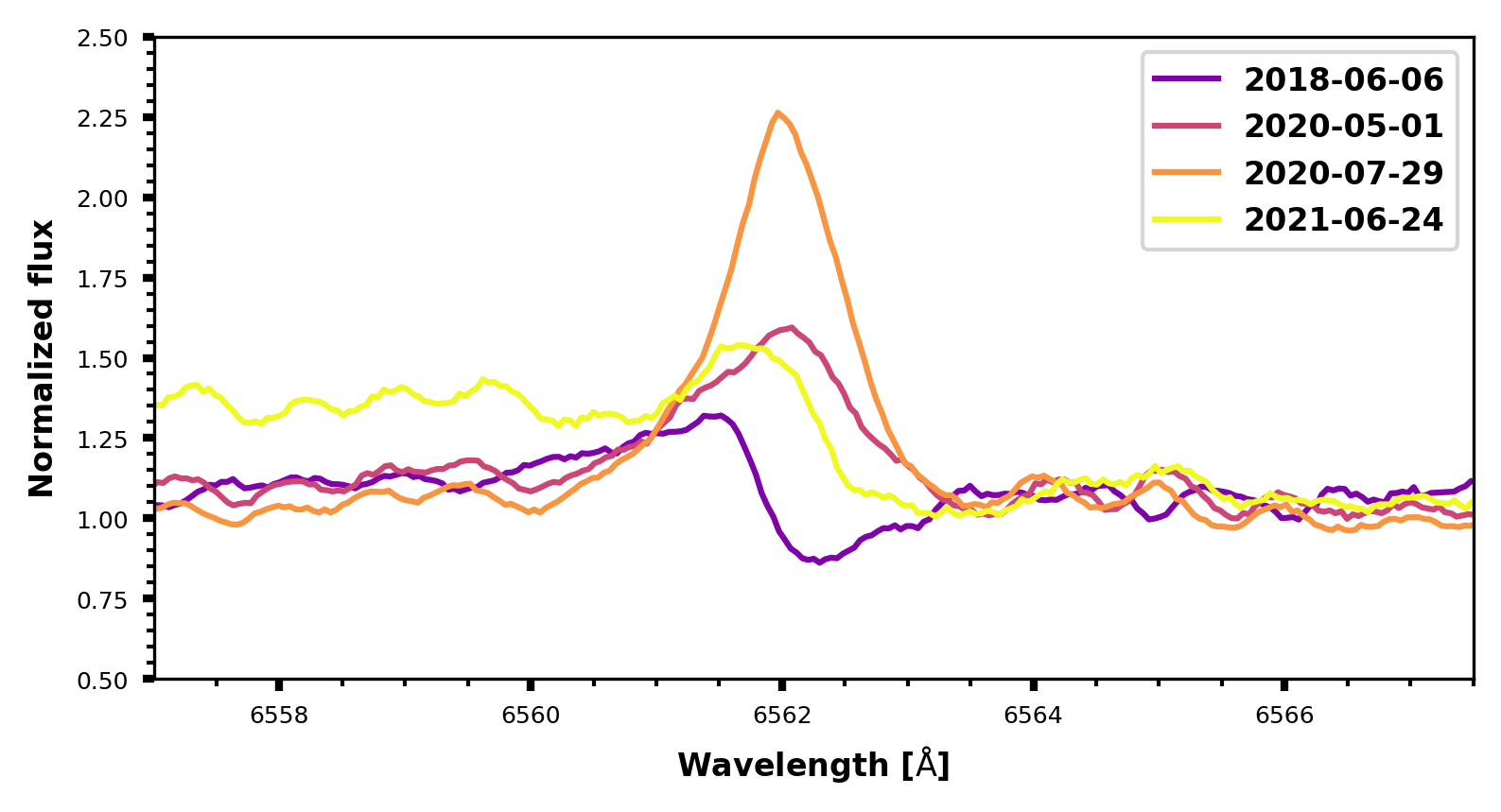}
    
    \caption{
    {\em Upper panel:}
    Radial velocity measurements and FWHM of CCF for VX Sgr from the STELLA telescope compared to $V$ and visual light curves from AAVSO. The data show similar behavior to that of Betelgeuse \citep{kravchenko21, jadlovsky24}, namely, the double-shock feature in the FWHM plot before the likely mass-loss event.
    {\em Lower panel:}
    H$\alpha$ for selected STELLA spectra showing the emergence of the emission component during the 2020-2021 event.}
    \label{fig:vx_sgr_rv}
\end{figure}

\begin{figure}[htpb]
    \includegraphics[width=0.5\textwidth, keepaspectratio]{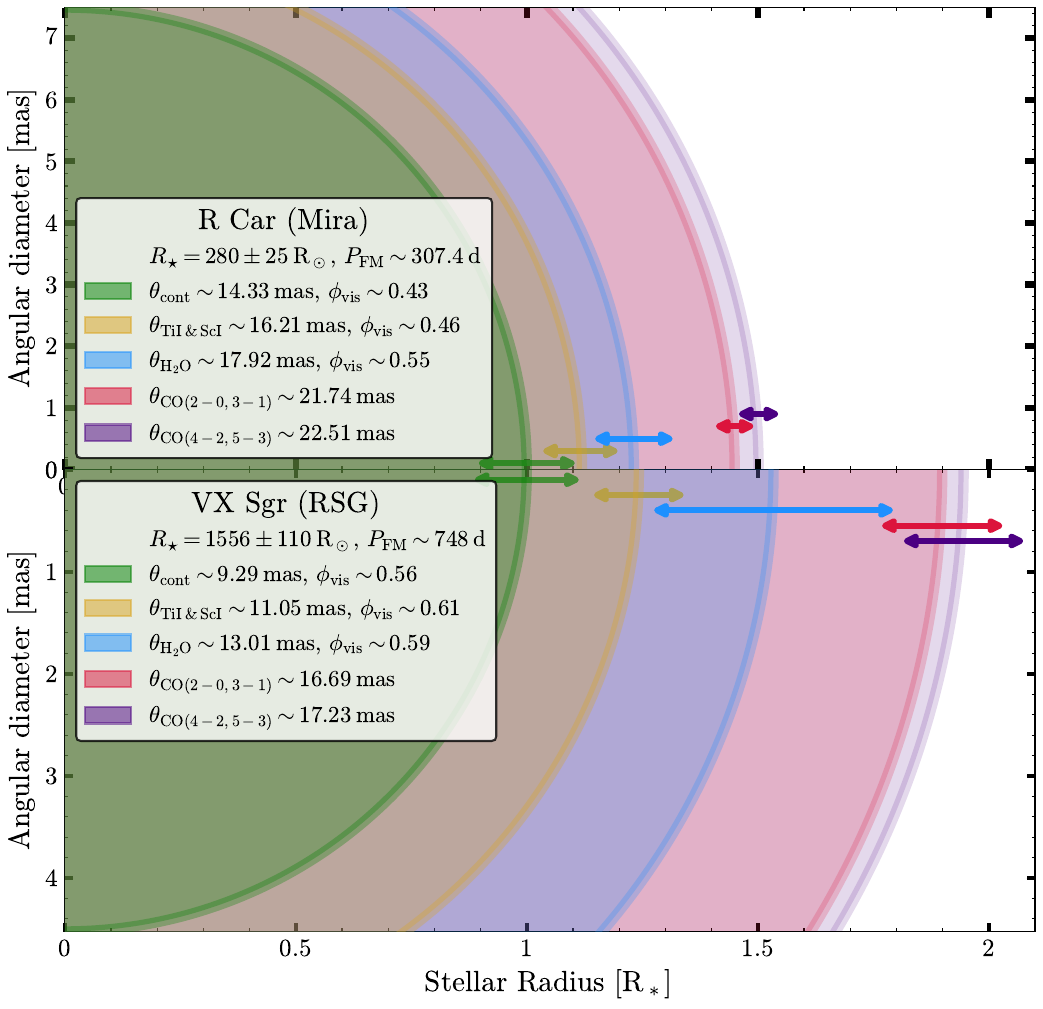}
    \caption{Schematic overview of our results for R Car and VX Sgr showing average atmospheric extension and diameter variations (arrows) scaled by the continuum (photopsheric) radius $\theta_{\rm cont}$. For VX Sgr, the properties based on the active cycle are shown.
    }
    \label{fig:schematic}
\end{figure}

\section{Conclusions}
\label{chapter:conclusions}
In this work, we have analyzed the largest sample of time series VLTI-GRAVITY observations up to the present time for two cool evolved stars: the Mira-type AGB star R Car and the RSG VX Sgr. A schematic overview of our results is shown in Fig. \ref{fig:schematic}. We studied the variability of the photosphere and atmospheric extension spanning several pulsation cycles and analyzed its relation to brightness variability. We estimated the angular diameters using a UD model. Our results demonstrate that the photosphere and inner atmospheric extension of AGB and RSG stars follow the brightness variability with a phase shift, showing maximum diameters between $\phi_{\rm vis} \sim 0.4$ and $0.6$. In other words, they are roughly anticorrelated with the visual light curve. At outer layers (up to $\sim 2\:  R_{\star}$), the variability becomes increasingly irregular, and the timescales are longer. Such behavior is compatible with the results published based on AGB RHD 3D simulations with CO5BOLD \citep[e.g.,][]{freytag23, freytag24} and was also predicted by earlier 1D CODEX models based on self-excited pulsation models \citep{ireland08, ireland11}.

For the Mira-type AGB star R Car with regular variability, we found that our determined angular diameters of photospheric and near-photospheric layers are phase-shifted to the brightness variability during each pulsation cycle. Maximum diameters are reached the earliest for the photosphere, at $\sim0.43$ (continuum), followed by $\ion{Ti}{i}$ and $ \ion{Sc}{i}$ at $\sim 0.46$ and H$_2$O at $\sim 0.55$, corresponding to a delay of about 2-3 weeks between the neighboring layers. The overall atmospheric extension reaches about $\sim 1.3-1.7 \: R_{\star}$ at the CO layers. In addition, R Car shows Br$\gamma$ emission in the flux near brightness maxima. For RSG VX Sgr, there is a similar behavior. However, due to its irregular nature, this RSG exhibits an extreme change to its pulsation properties during the active and quiescent cycles, with large-amplitude FM pulsations in the former and low-amplitude O1 pulsations in the latter. In the active cycle, the near-photospheric layers behave similarly to those in R Car and show maximum diameters between $ \phi_{\rm vis} \sim 0.57$ and $0.61$. However, here the phase shifts translate to a larger time delay of about 4-5 weeks. 
The H$_2$O bands show variations similar to those of the light curve as well, but with a large amplitude. The mean extension of CO reaches up to $\sim 1.5-2.0 \: R_{\star}$. 

Overall, R Car showed a mean photospheric diameter of $\theta_{\rm cont} = 14.33 {\pm0.11} \rm \: mas $, which translates to an estimated physical size of the photosphere of $R_{\star} = 280 \pm 25 \: \rm R_\odot$, with an FM pulsation amplitude of $\Delta R  \sim  35 \: \rm R_\odot $ ($\sim13 \%$ of $R_{\star} $). On the other hand, during the active cycle, VX Sgr showed $\theta_{\rm cont, active} = 9.29 {\pm0.10} \rm \: mas $, translating to $R_{\star, \rm active} = 1556 \pm 110 \: \rm R_\odot$, with a pulsation amplitude of $\Delta R  \sim  197 \: \rm R_\odot $ ($\sim13 \%$ of $R_{\star} $) and corresponding to pulsations in the FM. Meanwhile, during the quiescent cycle, it showed $\theta_{\rm cont, \rm quiescent} = 8.69 {\pm0.09} \rm \: mas $, yielding a smaller diameter, $R_{\star, \rm quiescent}= 1456 \pm 108 \: \rm R_\odot$, with a much smaller pulsation amplitude of $\Delta R  \sim  60 \: \rm R_\odot $ ($\sim4 \%$ of $R_{\star} $), likely corresponding to the O1.

Furthermore, at the end of its active cycle during 2020-2021, RSG VX Sgr likely showed an extreme mass-loss event preceded by two strong shocks. Following the event, the mode of pulsations changed from $\sim 748 \: \rm d$ to $\sim 316 \: \rm d$. This behavior is similar to that of Betelgeuse \citep{kravchenko21, dupree22, jadlovsky24}, which also showed two shocks preceding the Great Dimming. At the end of the active cycle in mid-2021, the molecular layers of H$_2$O and CO showed an extreme expansion, both reaching up to about $\sim 2.2 \:  R_{\star}$. Uniquely, during this event, VX Sgr showed strong Balmer emission, while Br$\gamma$ was also detected in the interferometric data. The hydrogen emission is a signature of a strong shock passing through the photosphere.

While both R Car and VX Sgr show extended atmospheres, which can appear observationally similar at single epochs, their time variability is quite different. R Car shows a regular variability of the angular radii of the photosphere, near-photospheric atomic and molecular layers, and more extended water vapor layers over several pulsation cycles. VX Sgr shows a more irregular variability of the angular radii of these layers over a similar time span, with an extreme event of levitation of these layers and a change in pulsation period. This supports the episodic occurrence of extreme shock and levitation events as a main driver of the mass-loss process in RSGs, as suggested by \citet{dupree22, humphreys22}, instead of or in addition to the regular pulsation observed in AGB stars. Furthermore, based on the luminosity of VX Sgr and its refined maser distance by \citet{xu18}, we compared our results to pulsation models \citep[e.g.,][]{joyce20}, and we conclude that VX Sgr is very likely an extremely massive RSG as opposed to a super-AGB star, as proposed in previous studies \citep[e.g.,][]{tabernero21}.

We also compared our time series of R Car to CO5BOLD simulations of AGB stars \citep{freytag12}, namely AGB model st28gm05n057 \citep{ahmad25}. We calculated the $K$-band intensity during several pulsation cycles. We found a qualitative agreement for the variability of the atmospheric extension, with the continuum and $\ion{Ti}{i}$ and $ \ion{Sc}{i}$ showing pulsations related to the light curve, although showing maximum angular diameters significantly preceding the minimum brightness. The AGB model is less extended than our observed stars, especially at the molecular bands of H$_2$O, while CO shows a reasonable agreement. Overall, the simulation shows pulsations as well as several non-spherical outflows, which are different in each pulsation cycle. The discrepancy in molecular bands of H$_2$O may be due to the gray Rosseland opacity approximation in the models \citep{hofner19}. This shows the necessity of properly accounting for wavelength-dependent opacity, as already done by \citet{freytag23}, to achieve better agreement with the observations. A radiative pressure on molecular lines could also play a major role \citep[e.g.,][]{josselin07}. 

Overall, we have demonstrated that new 3D models show a better agreement with the atmospheric extension than in previous studies \citep[e.g.,][]{arroyo15} and that it is possible to reproduce similar pulsation properties over several pulsation cycles. Future studies will compare the atmospheric extension of new AGB and RSG models in CO5BOLD, and other new 3D grids are also being developed, such as AREPO-RSG \citep{ma25}.

\begin{acknowledgements}
    We thank the anonymous referee for useful comments that improved the quality of the paper.
     
    We acknowledge R. Dorda and I. Negueruela for sharing STELLA spectra for VX Sgr from 2016 to 2021. Fruitful discussions with L. Molnár are also acknowledged. 
    
    DJ acknowledges support from the ESO-MEYS Training Programme of the public and from the ESO Studentship.

    DJ and JK were partly supported by grant GA \v{C}R 25-15910S.

    SH and BF acknowledge funding from the European Research Council (ERC) under the European Union’s Horizon 2020 research and innovation programme (Grant agreement No. 883867, project EXWINGS) and the Swedish Research Council (Vetenskapsradet, grant number 2019-04059).

    The computation of the CO5BOLD model was enabled by resources provided by the Swedish National Infrastructure for Computing (SNIC) and the National Academic Infrastructure for Supercomputing in Sweden (NAISS), partially funded by the Swedish Research Council through grant agreements no.~2018-05973 and no.~2022-06725.

     Based on observations made with the Very Large Telescope Interferometer (VLTI) at the Paranal Observatory of European Southern Observatory (ESO) under programs ID 0100.D-0835(C, F), 0101.D-0616(A), 0102.D-0197(A), 0103.D-0245(A, B, C), 105.207Y(002, 003, 004) and 115.27VK.001.

     Based on data obtained with the STELLA robotic telescopes in Tenerife, an AIP facility jointly operated by AIP and IAC.

     We acknowledge with thanks the variable star observations from the AAVSO International Database contributed by observers worldwide and used in this research.

\end{acknowledgements}

%
%
\bibliographystyle{aa} 
\bibliography{bibliography} 

%

\begin{appendix}
\onecolumn

\section{Observing logs} 
\begin{table*}[htbp]
        \centering
        \caption{Observations VX Sgr taken with VLTI-GRAVITY at the small VLTI-AT configuration. 
        } 
        \label{table:table_vxsgr} 
        \setlength{\extrarowheight}{3pt}
        \begin{tabular}{ccccccc}
\hline\hline
\text{Date} & \text{Time [UT]} & \text{Science exposure [s]}  & \text{Seeing ['']} & \text{Coherence time [ms]}  &  \text{Calibrators} & Epoch
\\
\hline
2018-04-09 & 07:37:30 & 10.0 & 0.39 & 8.88 & 24 Sgr, 28 Sgr & 1 \\ 
2019-03-26 & 08:38:02 & 10.0 & 0.89 & 5.79 & 24 Sgr, HD 176124 & 2 \\ 
2019-03-27 & 08:53:22 & 10.0 & 1.38 & 2.32 & 24 Sgr, HD 176124 & 2\\ 
2019-06-01 & 06:23:50 & 5.0 & 1.06 & 2.21 & 14 Sgr & 3 \\ 
2019-06-01 & 08:07:00 & 5.0 & 1.31 & 2.53 & 14 Sgr & 3 \\ 
2019-06-01 & 08:56:57 & 5.0 & 1.44 & 3.11 & 14 Sgr & 3\\ 
2019-06-04 & 02:36:50 & 5.0 & 0.94 & 5.58 & 14 Sgr & 3 \\ 
2019-07-31 & 00:24:19 & 5.0 & 0.97 & 3.03 & 14 Sgr & 4 \\ 
2019-07-31 & 03:45:03 & 5.0 & 1.35 & 2.91 & 14 Sgr & 4 \\ 
2019-07-31 & 22:58:04 & 5.0 & 0.98 & 2.62 & 14 Sgr & 4\\ 
2019-08-01 & 02:18:55 & 5.0 & 1.32 & 1.90 & 14 Sgr & 4\\ 
2019-08-01 & 02:35:28 & 5.0 & 1.40 & 1.93 & 14 Sgr & 4 \\ 
2019-08-02 & 03:32:50 & 5.0 & 1.78 & 1.56 & 14 Sgr & 4\\ 
2019-08-03 & 23:41:41 & 5.0 & 0.63 & 5.63 & 14 Sgr  & 4 \\ 
2019-09-04 & 02:35:58 & 5.0 & 0.54 & 3.84 & 14 Sgr & 5\\ 
2020-03-20 & 07:48:18 & 5.0 & 0.98 & 5.44 & 14 Sgr & 6 \\ 
2020-03-20 & 08:10:54 & 5.0 & 0.84 & 6.80 & 14 Sgr & 6\\ 
2021-08-30 & 02:24:52 & 0.3 & 1.29 & 1.68 & 14 Sgr & 7 \\ 
2021-08-31 & 01:35:48 & 0.3 & 0.84 & 2.84 & 14 Sgr  & 7 \\ 
2021-08-31 & 23:47:58 & 3.0 & 0.76 & 4.42 & 24 Sgr, HD 176124 & 7 \\ 
2022-04-30 & 09:02:21 & 3.0 & 0.56 & 6.02 & 24 Sgr, HD 176124 & 8 \\ 
2022-06-24 & 06:26:41 & 3.0 & 0.90 & 2.10 & 24 Sgr, HD 176124 & 9\\ 
2022-08-21 & 03:02:24 & 3.0 & 0.79 & 4.03 & 24 Sgr, HD 176124  & 10 \\ 
2022-08-21 & 03:14:43 & 3.0 & 0.9 & 3.26 & 24 Sgr, HD 176124 & 10 \\ 
2022-09-23 & 00:44:09 & 3.0 & 1.04 & 2.86 & 24 Sgr, HD 176124 & 11 \\ 
2022-10-14 & 00:24:04 & 3.0 & 1.04 & 6.24 & 24 Sgr, HD 176124 & 12\\ 
2022-10-14 & 00:45:02 & 3.0 & 0.68 & 5.47 & 24 Sgr, HD 176124 & 12 \\ 
2023-03-27 & 07:58:35 & 3.0 & 0.44 & 6.48 & 24 Sgr, HD 176124  & 13\\ 
2023-05-13 & 04:36:34 & 3.0 & 1.20 & 4.36 & 24 Sgr, HD 176124 & 14 \\ 
2023-06-08 & 07:01:38 & 3.0 & 0.38 & 11.10 & 24 Sgr, HD 176124 & 15 \\ 
2023-07-12 & 05:11:47 & 3.0 & 1.04 & 1.92 & 24 Sgr, HD 176124 & 16 \\ 
2023-08-10 & 03:41:39 & 3.0 & 1.27 & 1.52 & 24 Sgr, HD 176124 & 17\\ 
2023-09-22 & 01:41:55 & 3.0 & 1.38 & 2.28 & 24 Sgr, HD 176124 & 18 \\ 

2025-04-05 & 06:59:49 & 3.0 & 0.52 & 7.61 & 24 Sgr, HD 152636  & 19 \\ 
2025-05-19 & 09:02:23 & 3.0 & 0.65 & 6.01 & 24 Sgr, HD 152636  & 20 \\ 
2025-06-24 & 06:02:35  & 3.0 & 0.70  & 2.99 & 24 Sgr, HD 152636  & 21 \\ 
        \hline
        \end{tabular}

\end{table*}

\begin{table*}[htbp]
        \centering
        \caption{Same as Table \ref{table:table_vxsgr} but for R Car. 
        } 
        \label{table:table_rcar} 
        \setlength{\extrarowheight}{3pt}
        \begin{tabular}{ccccccc}
\hline\hline
\text{Date} & \text{Time [UT]} & \text{Science exposure [s]}  & \text{Seeing ['']} & \text{Coherence time [ms]}  & \text{Calibrators}  & \text{Epoch}  
\\
\hline
2018-01-27 & 05:19:03 & 3.0 & 0.70 & 5.34 & $\iota$ Car  & 1 \\ 
2018-01-27 & 06:15:11 & 3.0 & 0.81 & 5.22 & $\iota$ Car & 1 \\ 
2018-01-27 & 07:08:47 & 3.0 & 0.73 & 5.92 & $\iota$ Car & 1 \\ 
2018-01-27 & 08:13:16 & 3.0 & 0.50 & 6.25 & $\iota$ Car & 1 \\ 
2018-02-22 & 05:29:51 & 3.0 & 0.47 & 12.54 & $\iota$ Car & 2 \\ 
2018-02-25 & 05:27:26 & 3.0 & 0.33 & 13.34 & $\iota$ Car & 2 \\ 
2018-03-10 & 05:05:22 & 3.0 & 0.45 & 7.07 & $\iota$ Car & 2 \\ 
2018-03-12 & 00:44:15 & 3.0 & 0.79 & 4.90 & $\iota$ Car & 2 \\ 
2020-12-27 & 06:12:44 & 3.0 & 1.27 & 2.61 & $\beta$ Vol, t$^2$ Car & 3 \\ 
2021-02-21 & 06:15:38 & 3.0 & 0.54 & 9.12 & $\beta$ Vol, t$^2$ Car & 4 \\ 
2021-03-14 & 03:16:43 & 3.0 & 0.64 & 7.21 & $\beta$ Vol, t$^2$ Car & 5 \\ 
2021-05-20 & 23:25:02 & 3.0 & 0.92 & 2.67 & $\beta$ Vol, t$^2$ Car & 6 \\ 
2022-02-03 & 06:58:03 & 3.0 & 0.68 & 8.31 & t$^2$ Car  & 7 \\ 
2022-04-28 & 01:28:19 & 3.0 & 0.71 & 3.24 & $\beta$ Vol, t$^2$ Car & 8 \\ 
2022-11-20 & 07:15:40 & 3.0 & 0.61 & 4.99 & $\beta$ Vol, t$^2$ Car  & 9 \\ 
2023-01-14 & 05:31:04 & 3.0 & 1.17 & 3.43 & $\beta$ Vol, t$^2$ Car & 10 \\ 
2023-02-25 & 01:06:22 & 3.0 & 0.78 & 4.02 & $\beta$ Vol, t$^2$ Car & 11 \\ 
2023-05-13 & 00:20:31 & 3.0 & 0.55 & 8.22 & $\beta$ Vol, t$^2$ Car & 12 \\ 
        \hline
        \end{tabular}
\end{table*}

\begin{table*}[htbp]
        \centering
        \caption{Properties of the calibrators used for VLTI-GRAVITY observations of R Car and VX Sgr.} 
        \label{table:table_cals} 
        \setlength{\extrarowheight}{3pt}
        \begin{tabular}{cccccc}
\hline\hline
\text{Calibrator} & \text{Spectral Type}  & \text{LDD [mas]} & \text{UDK [mas]}  & $T_{\rm eff} \: \rm [K] $ & $\log g$ (cgs) 
\\
\hline
14 Sgr & K2III & 2.28 $\pm$ 0.23 & 2.23 & 3900 & 1.5 \\ 
24 Sgr & K3III & 3.52 $\pm$ 0.36 & 3.44 & 4000 & 0.5 \\ 
28 Sgr & K5III & 2.81$\pm$ 0.25 & 2.74 & 3900 & 1.5 \\ 
$\beta$ Vol & K2III & 2.95$\pm$ 0.30 & 2.88 & 4750 & 3 \\ 
$\iota$ Car & A7Ib & 1.87 $\pm$ 0.19 & 1.84 & 8000 & 2.0 \\ 
t$^2$ Car & K4/5(III) & 3.67 $\pm$ 0.40 & 3.58 & 3900 & 1.5 \\ 
HD 176124 & M3III & 4.25 $\pm$ 0.44 & 4.21 & 3300 & 2.0 \\ 
HD 152636 & K5III & 2.39 $\pm$ 0.23 & 2.33 & 3800 & 2.0 \\ 

        \hline
        \end{tabular}
\tablefoot{Listed are the limb-darkened disk (LDD) and K-band diameters (UDK) from \citet{bourges17}. We also list properties of MARCS models \citep{gustaf08} used for flux calibration, with $T_{\rm eff} $ and $\log g$ estimated from spectral type.}
\end{table*}

\FloatBarrier 
\onecolumn

\begin{figure*}[htbp]
\section{Full VLTI dataset of R Car} 
\label{appendix:r_car_data}
    \includegraphics[width=1\textwidth, keepaspectratio]{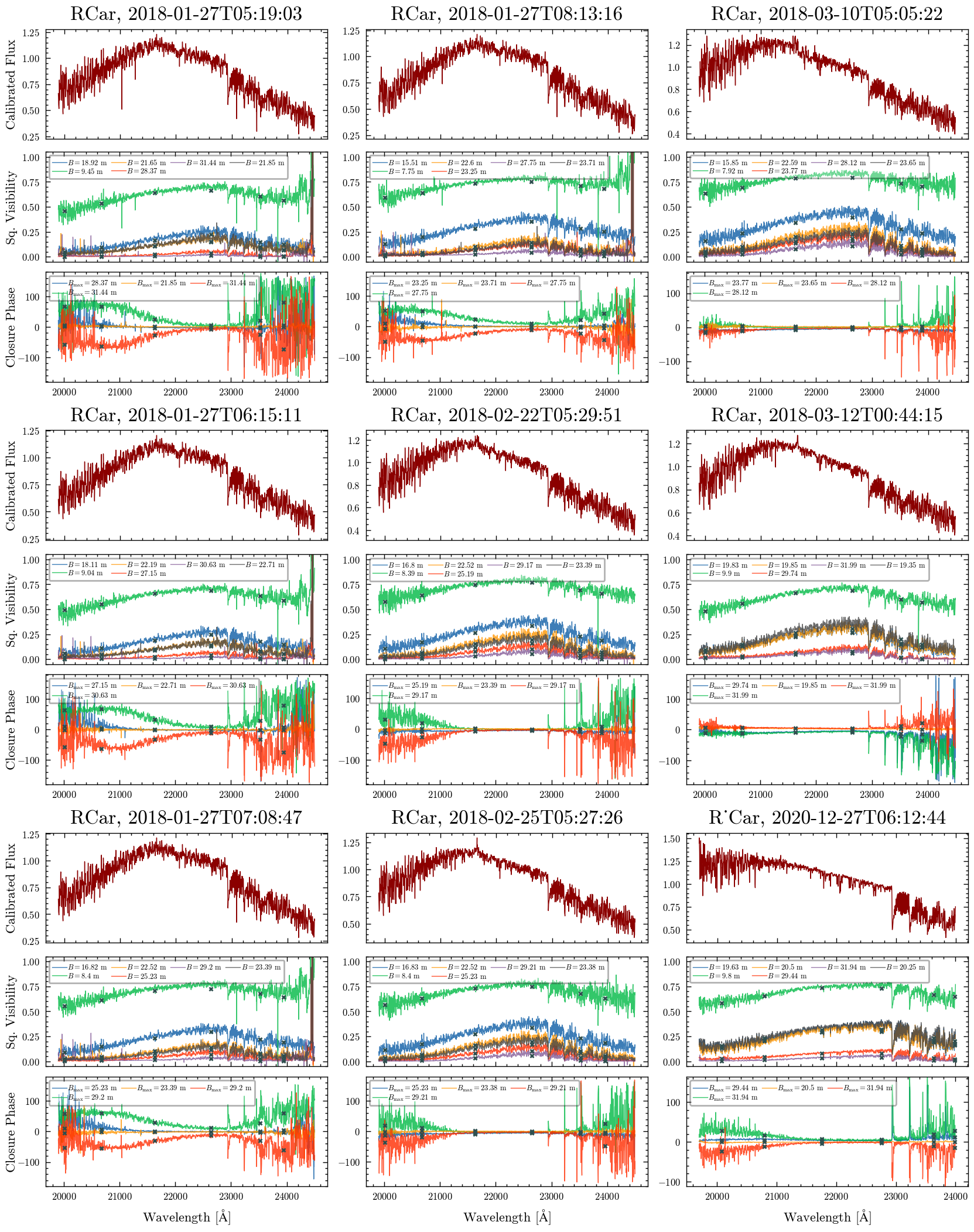}
    \caption{Full spectro-interferometric dataset for R Car. Namely flux (calibrated using the spectral transfer function), squared visibility, and closure phase. Data from the fringe tracker are also shown as black crosses. }
    \label{fig:r_car_plots_1}
\end{figure*}

\begin{figure*}[htbp]
\addtocounter{figure}{-1}
    \includegraphics[width=1\textwidth, keepaspectratio]{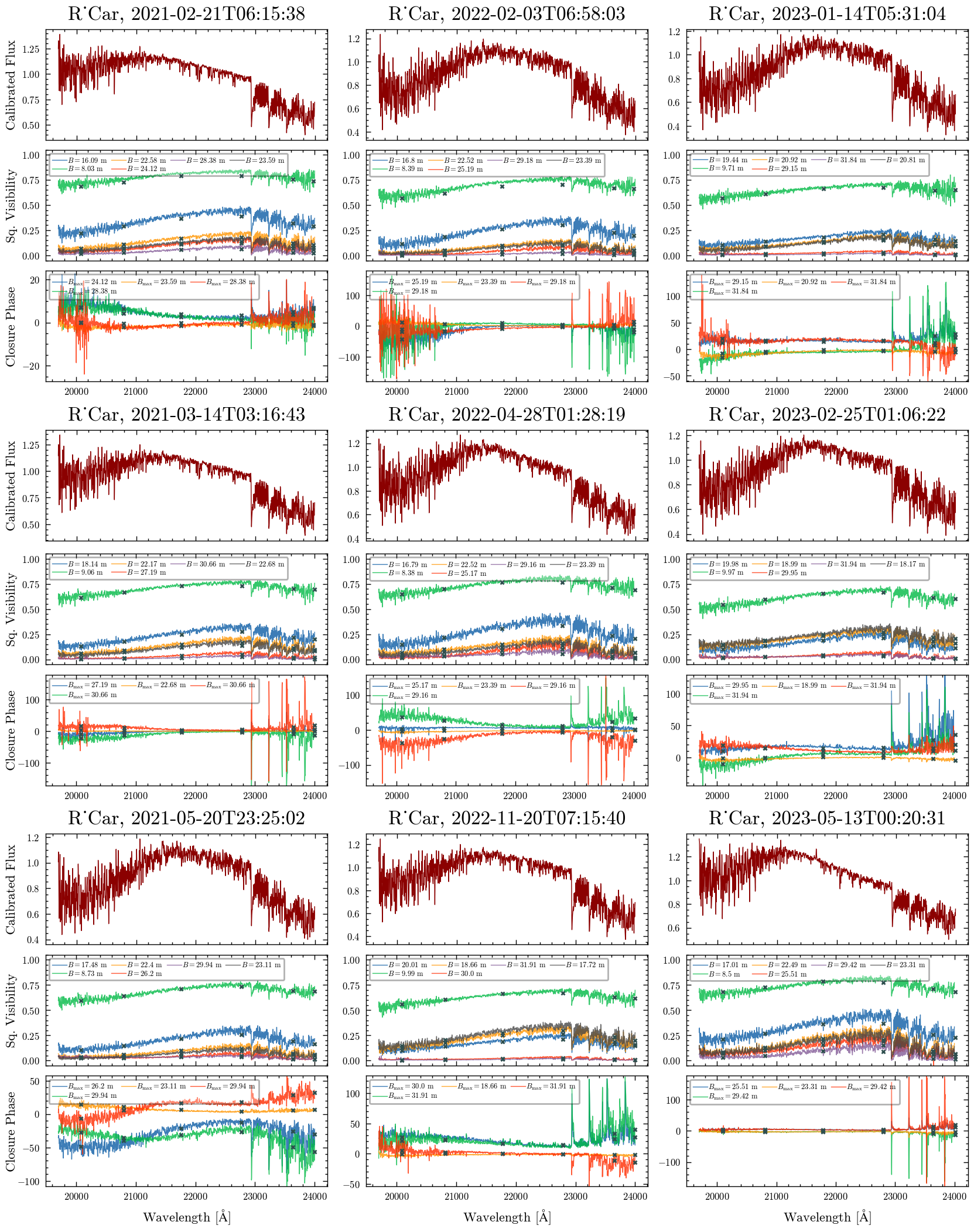}
    \caption{Continued.}
    \label{fig:r_car_plots_2}
\end{figure*}

\begin{figure*}[htbp]
\section{Full VLTI dataset of VX Sgr} 
\label{appendix:vx_sgr_data}
    \includegraphics[width=1\textwidth, keepaspectratio]{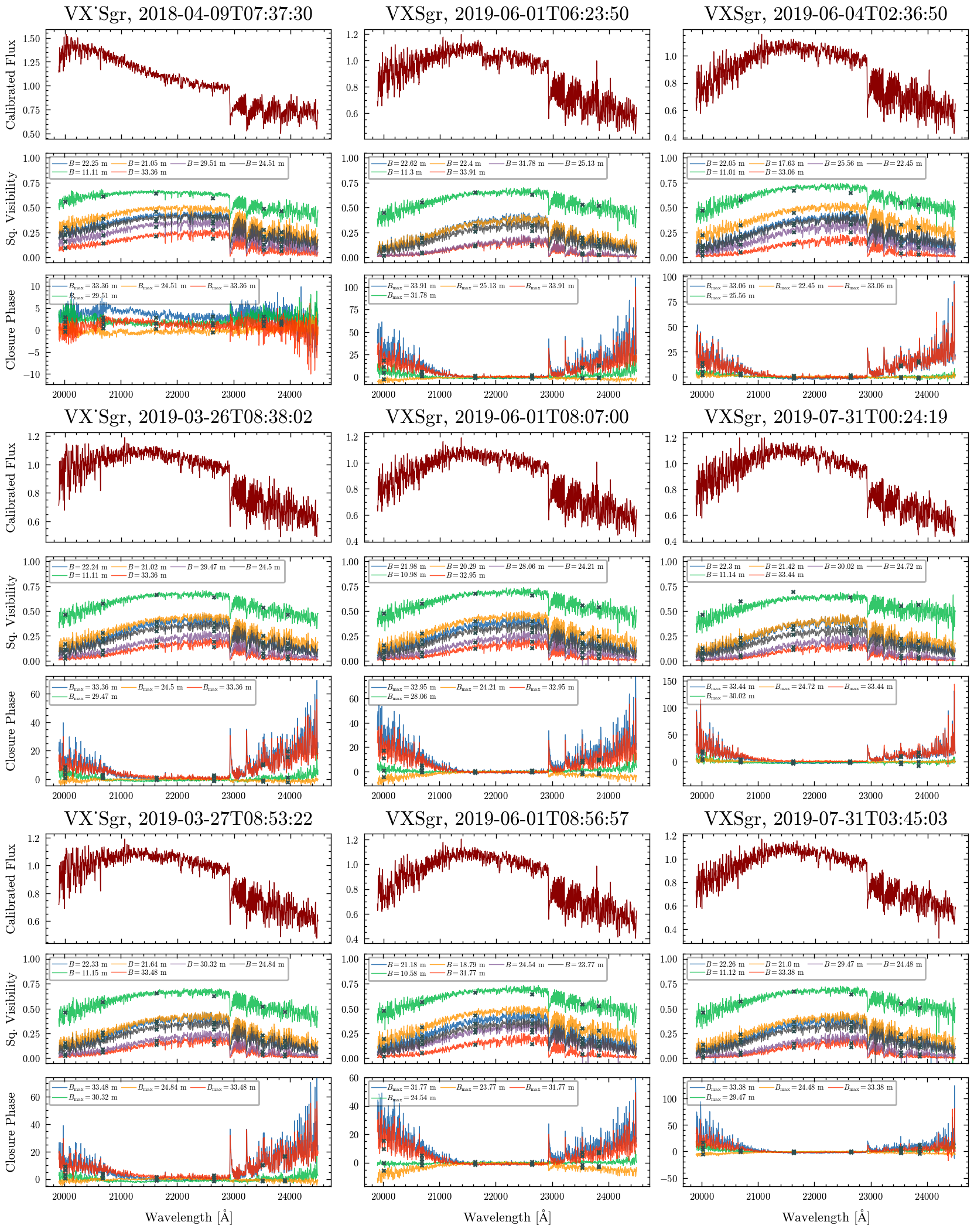}
    \caption{Same as Fig.~\ref{fig:r_car_plots_1} but for VX Sgr.}
    \label{fig:vx_sgr_plots_1}
\end{figure*}

\begin{figure*}[htbp]
\addtocounter{figure}{-1}
    \includegraphics[width=1\textwidth, keepaspectratio]{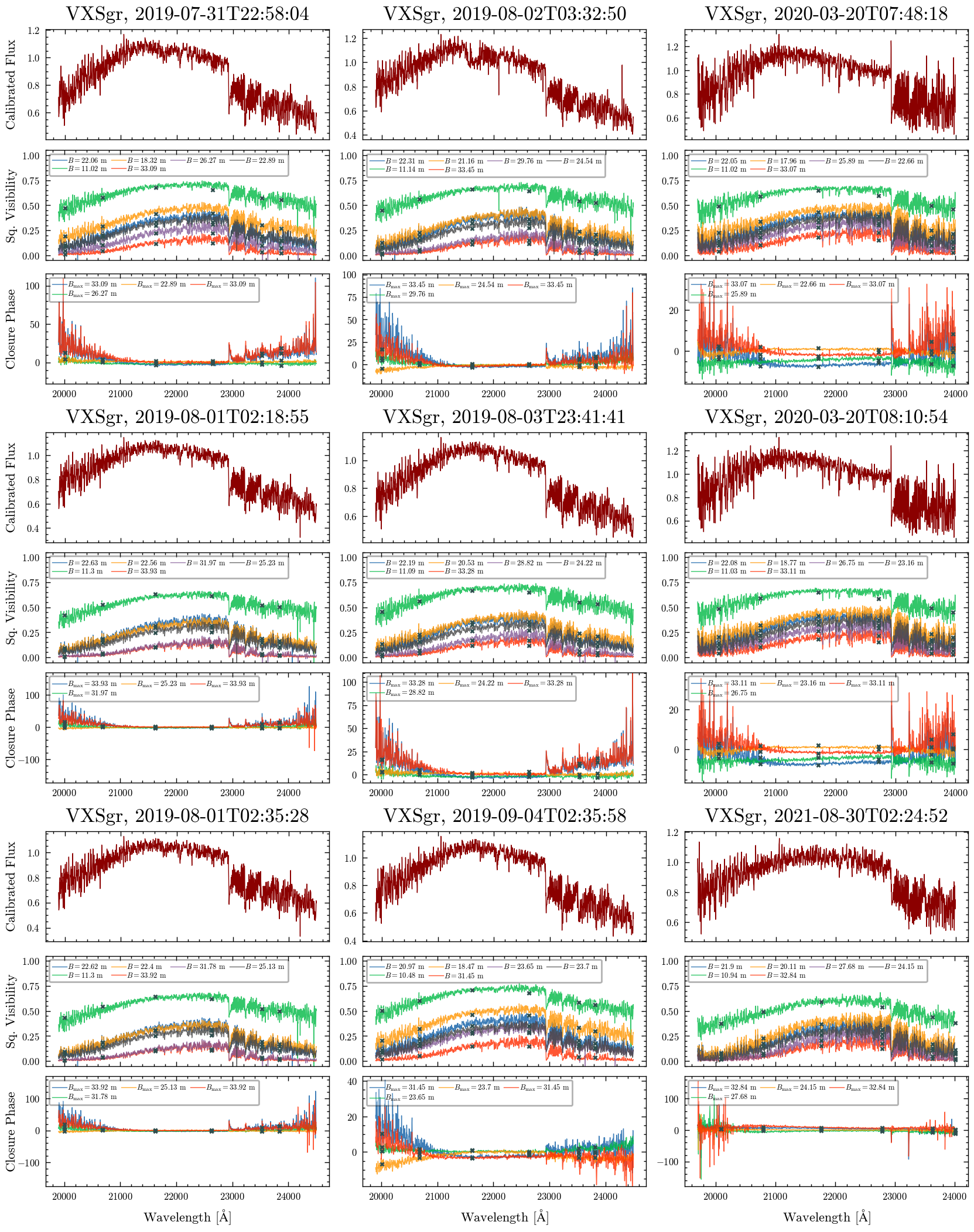}
    \caption{Continued.}
    \label{fig:vx_sgr_plots_2}
\end{figure*}

\begin{figure*}[htbp]
\addtocounter{figure}{-1}
    \includegraphics[width=1\textwidth, keepaspectratio]{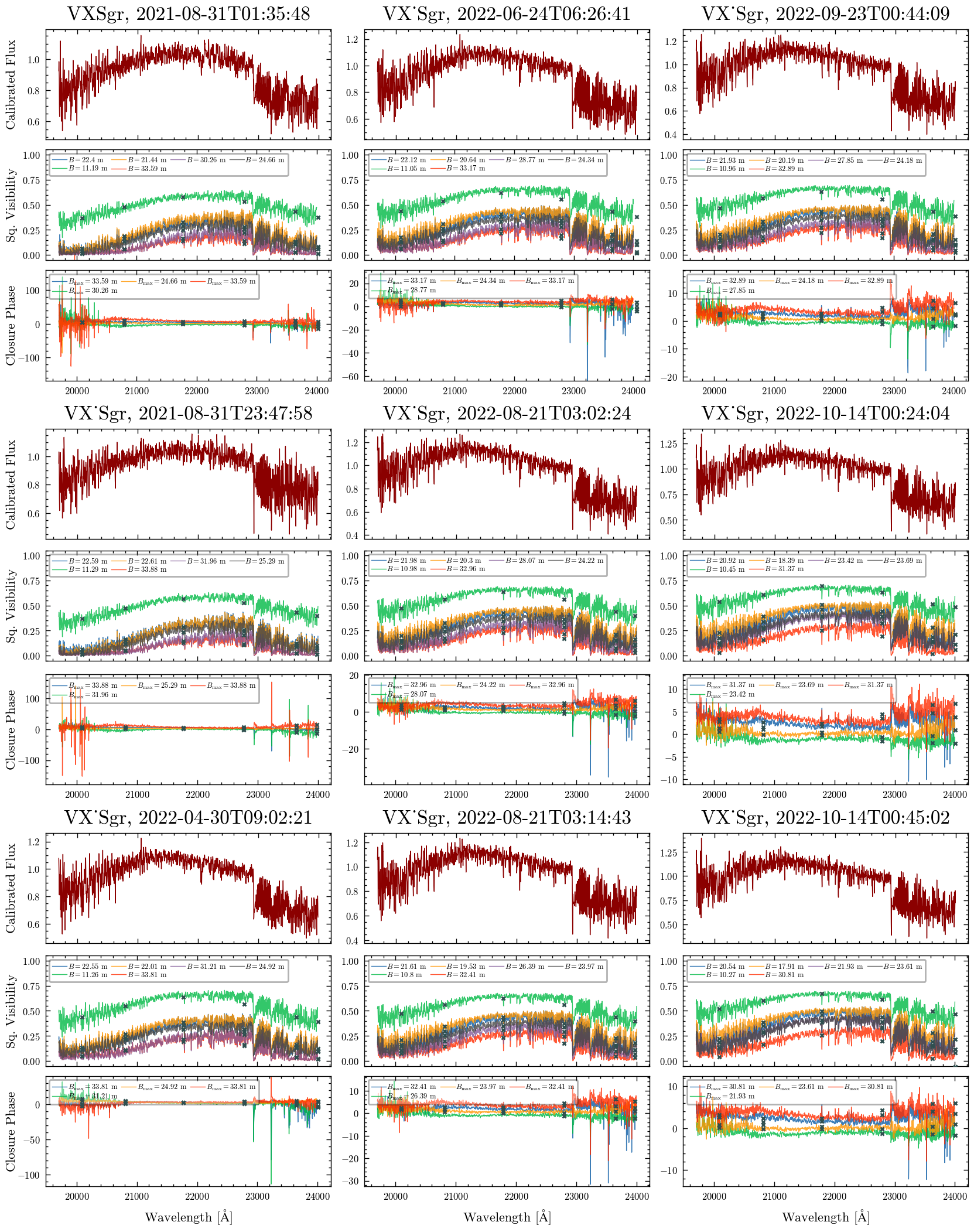}
    \caption{Continued.}
    \label{fig:vx_sgr_plots_3}
\end{figure*}

\begin{figure*}[htbp]
\addtocounter{figure}{-1}
    \includegraphics[width=1\textwidth, keepaspectratio]{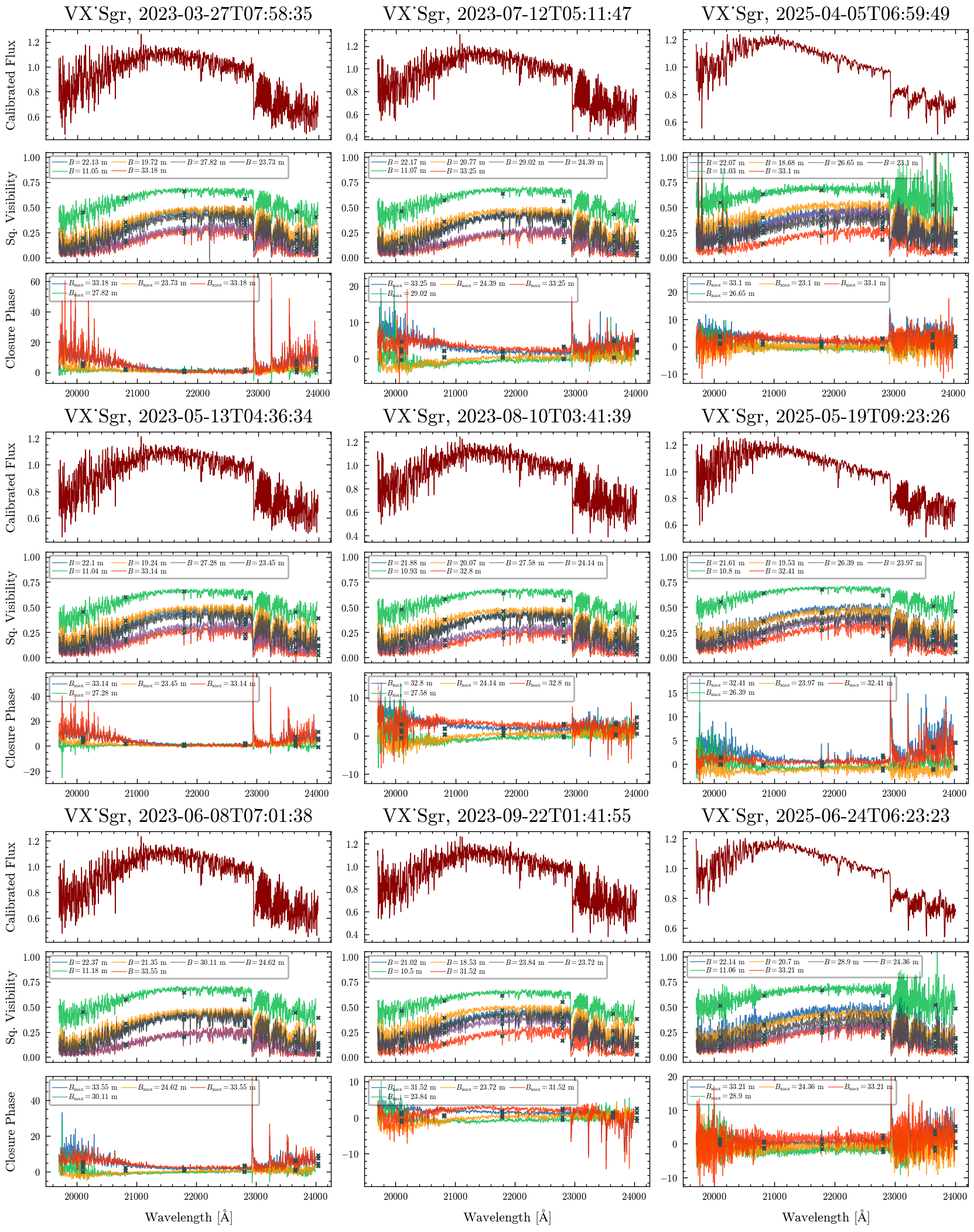}
    \caption{Continued.}
    \label{fig:vx_sgr_plots_4}
\end{figure*}

\end{appendix}

\end{document}